\documentclass[twocolumn,letterpaper,10pt]{article}
\usepackage{usenix2019}
\usepackage{epsfig}
\usepackage{graphicx}
\usepackage{color}
\usepackage{amsmath}
\usepackage[T1]{fontenc}
\usepackage{fixltx2e}
\usepackage{hyperref}
\usepackage{breakurl}
\usepackage{flushend}
\usepackage{tabularx}
\usepackage[normalem]{ulem}
\usepackage{textcomp}
\usepackage{multirow}
\usepackage{times}
\usepackage{courier}
\usepackage{listings}
\usepackage{relsize}
\usepackage{subcaption}
\newcommand{\erim}{ERIM}
\newcommand{\tlib}{$\mbox{T}$}
\newcommand{\uapp}{$\mbox{U}$}
\newcommand{\tmem}{$\mbox{M}_{\mbox{\smaller T}}$}

\newcommand{\umem}{$\mbox{M}_{\mbox{\smaller U}}$}

\newcommand{\todo}[1]{\textcolor{red}{TODO:{#1}}}

\usepackage[compact]{titlesec}
\usepackage{titling}
\begin{document}
\title{\Large \bf \textsc{\erim}: Secure, Efficient In-process
  Isolation with Protection Keys (MPK)}

\author{Anjo Vahldiek-Oberwagner \qquad Eslam Elnikety \qquad Nuno O. Duarte\\Michael Sammler \qquad Peter Druschel \qquad Deepak Garg\\
Max Planck Institute for Software Systems (MPI-SWS), Saarland Informatics Campus} 

\date{}
%\date{\vspace{-1.5cm}}

\maketitle

\pagestyle{empty}
\thispagestyle{empty}

\begin{abstract}
%\subsection*{Abstract}
Isolating sensitive state and data can increase the security and
robustness of many applications.  Examples include protecting
cryptographic keys against exploits like OpenSSL's Heartbleed bug or
protecting a language runtime from native libraries written in unsafe
languages. When runtime references across isolation boundaries occur
relatively infrequently, then conventional page-based hardware
isolation can be used, because the cost of kernel- or
hypervisor-mediated domain switching is tolerable. However, some
applications, such as the isolation of cryptographic session keys in
network-facing services, require very frequent domain switching. In
such applications, the overhead of kernel- or hypervisor-mediated
domain switching is prohibitive.

In this paper, we present {\erim}, a novel technique that provides
hardware-enforced isolation with low overhead on x86 CPUs, even at
high switching rates ({\erim}'s measured overhead is less than 1\% for
100,000 switches per second).  The key idea is to combine protection
keys (MPKs), a feature recently added to x86 that allows protection
domain switches in userspace, with binary inspection to prevent
circumvention. We show that ERIM can be applied with little effort to
new and existing applications, doesn't require compiler changes, can
run on a stock Linux kernel, and has low runtime overhead even at high
domain switching rates.

\end{abstract}

\section{Introduction}

It is good software security practice to partition sensitive data and
code into isolated components, thereby limiting the effects of bugs
and vulnerabilities in a component to the confidentiality and
integrity of that component's data. For instance, isolating
cryptographic keys in a network-facing service can thwart
vulnerabilities like the OpenSSL Heartbleed bug~\cite{heartbleed};
isolating a managed language's runtime can protect its security
invariants from bugs and vulnerabilities in co-linked native
libraries; and, isolating jump tables can prevent attacks on an
application's control flow.

%foremost requires \emph{memory isolation}, which

%\todo{Reviewer A: Compare with VMFUNC/MPX - next two paragraphs -
%  consider to add table of costs: runtime overhead | switch overhead
%  for MPX, VMFUNC, LwC, WebAssembly}
%\todo{Reviewer A: missing
%  citation Datashield~\cite{Carr:2017:DataShield}}

Isolation prevents an untrusted component from directly accessing the
private memory of other components.  Broadly speaking, isolation can
be enforced using one of two approaches. First, in software fault
isolation (SFI)~\cite{sfi}, one instruments the code of untrusted
components with bounds checks on indirect memory accesses, to prevent
access to other components' memory.  The bounds checks can be added by
the compiler or through binary rewriting.  Bounds checks impose
overhead on the execution of all untrusted components; additional
overhead is required to prevent control-flow
hijacks~\cite{Koning2017}, which could circumvent the bounds checks.
On x86-64, pointer masking-based SFI techniques like Native
Client~\cite{nacl:x64} incur overheads of up to 42\% on the execution
of untrusted code~\cite{Koning2017}. Even with hardware-supported
bounds checks, like those supported by the Intel MPX ISA
extension~\cite{intel-mpx}, the overhead is up to 30\%, as shown in by
Koning \emph{et al.}~\cite{Koning2017} and later in
Section~\ref{sec:eval:comparison}.
  
%  Similar results were shown in~\cite{Carr:2017:DataShield,Koning2017}.
  
%\todo{Reviewer A, C: Missing citations nested kernel + LOTR}
Another approach is to use hardware page protection for memory
isolation~\cite{Belay2012,Bittau2008,Chen2016,Lee:2018:LOTR,Litton2016,Liu2015}.
Here, access checks are performed in hardware as part of the address
translation with no additional overhead on execution {\em within} a
component.  However, transferring control {\em between} components
requires a switch to kernel or hypervisor mode in order to change the
(extended) page table base.
%
%% \footnote{Using x86 memory segmentation instead of page tables, as in
%%   Native Client~\cite{nativeclient}, avoids most of the switch
%%   cost. However, full support for segmentation is no longer available
%%   in the x86-64 and other modern architectures.}
%
Recent work such as Wedge, Shreds, SeCage, SMVs, and light-weight
contexts (lwCs)~\cite{Bittau2008,Chen2016,Hsu16ccs,Litton2016,Liu2015}
have reduced the overhead of such switching, but the cost is still
substantial. For instance, Litton \emph{et
    al.}~\cite{Litton2016} report a switching cost of about 1us per
  switch for lwCs, which use kernel-managed page tables for in-process
  isolation. This amounts to an overhead of nearly 10\% for an
  application that switches 100,000 times a second and, in our
  experiments, an overhead of up to 65\% on the throughput of the web
  server NGINX when lwCs are used to isolate session keys
  (Section~\ref{sec:eval:comparison}).
%
%% In particular, if lwCs are used to isolate
%% \emph{session} keys in a web server like NGINX,%
%% %
%% \footnote{Isolating a web server's session keys is a challenging use
%%   case because session keys are accessed very frequently and their
%%   isolation results in very high switching rates. In contrast,
%%   long-term signing keys are accessed only infrequently and can be
%%   isolated with low overhead using many isolation techniques.}
%% %
%% the switch cost results in an overhead of up to 65\% on
%% the sustained throughput (see Section~\ref{sec:eval:comparison}).
%
Techniques based on Intel VT-x extended page tables with
VMFUNC~\cite{Liu2015} have less overhead, but the overhead is still
high---up to 14.4\% on NGINX's throughput in our experiments
(Section~\ref{sec:eval:comparison}).

%% Part of this overhead also comes from process-level virtualization,
%% where all syscalls are converted into hypercalls, which can be up to
%% 2.4x more costly than syscalls, and extended page table walks, which
%% can be up to 6x more costly than ordinary page table walks. Hence,
%% applications that, unlike our web server, are either syscall-heavy or
%% have high TLB-miss rates will likely have even higher overheads.

%% NGINX is not syscall-heavy and does not cause high TLB-miss 
%% benchmark, other applications may experience higher overheads,
%% due to system call overhead of upto 6x and page walk overhead 
%% of upto 2.4x. 

%\updated{Table~\ref{tab:comparingisolation}
%  compares the efficiency of each approach.}
\if 0
\begin{table}[]
\begin{tabular}{l|p{2cm}|p{2cm}}
Technique & Runtime overhead & Switch Overhead \\\hline
MPX-based SFI~\cite{Carr:2017:DataShield, Koning2017} & 10s of \%  & none\\\hline
LwC~\cite{Litton2016} & none & 10\% \\\hline
VMFUNC~\cite{Liu2015} & none & 3-5\%   \\\hline
\end{tabular}
\caption{Overhead comparison of different memory isolation techniques. (Switch overhead in \% per 100,000 switches/s) \todo{Numbers are too vague?}}
\label{tab:comparingisolation}
\end{table}
\fi

In this paper, we present {\erim}, the first isolation technique for
x86 that combines near-zero overhead on in-component execution with
very low cost switching among components. {\erim} relies on a recent
x86 ISA extension called protection keys (MPK)~\cite{Intel2016}.  With
MPK, each virtual page can be tagged with a 4-bit domain id, thus
partitioning a process's address space into up to 16 disjoint
domains. A special register, PKRU, that is local to each logical core
determines which domains the core can read or write. Switching domain
permissions requires writing the PKRU register in userspace, which
takes only 11--260 cycles on current Intel CPUs, corresponding to an
overhead of 0.07\% to 1.0\% per 100,000 switches/s on a 2.6 GHz
CPU. This amounts to an overhead of at most 4.8\% on the throughput of
NGINX when isolating all session keys, which is up to 6.3x, 13.5x and
3x lower than the overhead of similar protection using SFI (with Intel
MPX), lwCs and Intel VT-x, respectively.

%\todo{Reviewer A: focus on a single matrix like \% overhead per 100,000 switches/s}

However, MPK by itself does not provide strong security because a
compromised or malicious component can simply write to the PKRU
register and grant itself permission to access any component. {\erim}
relies on \emph{binary inspection} to ensure that all occurrences of
instructions that update the PKRU in the binary are \emph{safe}, i.e.,
they cannot be exploited to gain unauthorized access. With this,
{\erim} provides isolation without requiring control-flow integrity in
untrusted code, and therefore avoids the runtime overhead of ensuring
control-flow integrity in unsafe languages.

While {\erim}'s binary inspection enforces the safety of its MPK-based
isolation, it creates a potential usability issue: What to do if a
binary has \emph{unintentional} occurrences of PKRU-updating
instructions?  Since x86 does not require instruction alignment, such
occurrences could arise within a longer instruction, or spanning the
bytes of two or more adjacent instructions. Any such sequence could be
exploited by a control-flow hijack attack and must be
rejected by the binary inspection mechanism. To handle such cases, we
describe a novel procedure to \emph{rewrite} any instruction sequence
containing an unaligned PKRU-updating instruction to a functionally
equivalent sequence without the instruction. This rewriting
procedure can be integrated with a compiler or our binary inspection.

{\erim} is the first technique that enables efficient isolation in
applications that require very high domain switching rates
(\textasciitilde$10^5$/s or more) and also spend significant time
executing inside untrusted components.  We evaluate our {\erim}
prototype on three such applications: 1) Isolating the frequently
accessed session keys in a web server (NGINX), 2) isolating a managed
language runtime from native libraries written in unsafe languages,
and 3) efficiently isolating the safe region in code-pointer
integrity~\cite{cpi}. In all cases, we observe switching rates of
order $10^5$ or more per second per core. {\erim} provides strong,
hardware-based isolation in all these cases, with overheads
  that are considerably lower than those of existing
  techniques. Moreover, {\erim} does not require compiler support and
can run on stock Linux.

In summary, this paper makes the following contributions. 1) We
present {\erim}, an efficient memory isolation technique that relies
on a combination of Intel's MPK ISA extension and binary inspection,
but does not require or assume control-flow integrity. 2) We describe
a complete rewriting procedure to ensure binaries cannot be exploited
to circumvent {\erim}. 3) We show that {\erim} can protect
applications with high inter-component switching rates with low
overhead, unlike techniques based on hardware (extended) page tables
and SFI (even with hardware support).

\section{Background and related work}
\label{erim:sec:background}
\label{erim:sec:relatedwork}

In this section, we survey background and related work.  Enforcing
relevant security or correctness invariants while trusting only a
small portion of an application's code generally requires {\em data
  encapsulation}. Encapsulation itself requires {\em isolating}
sensitive data so it cannot be accessed by untrusted code, and
facilitating \emph{switches} to trusted code that has access to the
isolated state.
%
%If the encapsulated code and data additionally functions as a {\em
%  reference monitor}~\cite{Anderson1972} that constrains the behavior
%of untrusted code, then a {\em mediation} technique may be required to
%ensure untrusted code cannot invoke sensitive operations without
%intervention.
%
%For instance, an application might wish to isolate its cryptographic
%keys and use SCT to invoke the crypto functions, which require access
%to the keys. If the application additionally wishes to prevent
%unencrypted communication, it must disallow or redirect I/O syscalls
%by untrusted code.
%
We survey techniques for isolation and switching provided by operating
systems, hypervisors, compilers, language runtimes, and binary
rewriting, as well as other work that uses MPK for memory isolation.

\paragraph{OS-based techniques}
Isolation can be easily achieved by placing application components in
separate OS processes.
%If required, mediation can be achieved using a capability sandbox
%(e.g., FreeBSD's Capsicum~\cite{Watson2010}) to control which syscalls
%a process can invoke.
However, this method has high overhead even with a moderate rate of
cross-component invocation. Novel kernel abstractions like
light-weight contexts (lwCs)~\cite{Litton2016}, secure memory views
(SMVs)~\cite{Hsu16ccs} and nested
kernels~\cite{Dautenhahn:2015:nestedkernel},
%\todo{Reviewer A: adding nested kernels}
combined with additional compiler support as in Shreds~\cite{Chen2016}
or runtime analysis tools as in Wedge~\cite{Bittau2008}, have reduced
the cost of such data encapsulation to the point where isolating
\emph{long-term} signing keys in a web server is feasible with little
overhead~\cite{Litton2016}. Settings that require more frequent
switches like isolating \emph{session keys} or the safe region in
CPI~\cite{cpi}, however, remain beyond the reach of OS-based techniques.

Mimosa~\cite{mimosa} relies on the Intel TSX hardware transactional
memory support to protect private cryptographic keys from software
vulnerabilities and cold-boot attacks. Mimosa restricts cleartext keys
to exist only within uncommitted transactions, and TSX ensures that an
uncommitted transaction's data is never written to the DRAM or other
cores. Unlike \erim, which is a general-purpose isolation technique,
Mimosa specifically targets cryptographic keys, and is constrained by
hardware capacity limits of TSX.

%% Private keys are stored
%% encrypted in memory. Accesses to private keys are performed within a
%% transaction that first decrypts the private key using a
%% register-backed master key, performs the access, wipes the cleartext
%% key, and then commits. The cleartext key never exists outside an
%% uncommitted transaction and TSX ensures that an uncommitted
%% transaction's data is never exposed to the DRAM or other cores. In
%% Mimosa, private key computations are constrained by the hardware
%% capacity limitations of TSX. In contrast, \erim\ places no
%% restrictions on the footprint of computations within secure
%% compartments.

\paragraph{Virtualization-based techniques}
In-process data encapsulation can be provided by a hypervisor.
Dune~\cite{Belay2012} enables user-level processes to implement
isolated compartments by leveraging the Intel VT-x x86 virtualization
ISA extensions~\cite{Intel2016}. Koning et al.~\cite{Koning2017}
sketch how to use the VT-x VMFUNC instruction to switch extended page
tables in order to achieve in-process data
isolation. SeCage~\cite{Liu2015} similarly relies on VMFUNC to switch
between isolated compartments. SeCage also provides static and dynamic
program analysis based techniques to automatically partition
monolithic software into compartments, which is orthogonal to our
work. TrustVisor~\cite{trustvisor} uses a thin hypervisor and nested
page tables to support isolation and additionally supports code
attestation.  SIM~\cite{Sharif09ccs} relies on VT-x to isolate a
security monitor within an untrusted guest VM, where it can access
guest memory with native speed.  In addition to the overhead of the
VMFUNC calls during switching, these techniques incur overheads on TLB
misses and syscalls due to the use of extended page tables and
hypercalls, respectively. Overall, the overheads of
virtualization-based encapsulation are much higher than those of
{\erim}.

Nexen~\cite{nexen} decomposes the Xen hypervisor into isolated
components and a security monitor, using page-based protection within
the hypervisor's privilege ring 0. Control of the MMU is restricted to
the monitor; compartments are de-privileged by scanning and removing
exploitable MMU-modifying instructions. The goal of Nexen is quite
different from \erim's: Nexen aims to isolate co-hosted VMs and the
hypervisor's components from each other, while \erim\ isolates
components of a user process. Like \erim\, Nexen scans for and removes
exploitable instructions.
%% However, by using MPK, \erim's isolation operates
%% entirely in ring 3 with only limited support from the OS kernel.

\paragraph{Language and runtime techniques}
Memory isolation can be provided as part of a memory-safe programming
language. This encapsulation is efficient if most of the checks can be
done statically. However, such isolation is language-specific, relies
on the compiler and runtime, and can be undermined by co-linked
libraries written in unsafe languages.

Software fault isolation (SFI)~\cite{sfi} provides memory isolation in
unsafe languages using runtime memory access checks inserted by the
compiler or by rewriting binaries. SFI imposes a continuous overhead
on the execution of untrusted code. Additionally, SFI by itself does
not protect against attacks that hijack control flow (to possibly
bypass the memory access checks). To get strong security, SFI must be
coupled with an additional technique for control-flow integrity
(CFI)~\cite{Abadi2005}. However, existing CFI solutions have
nontrivial overhead. For example, code-pointer integrity (CPI), one of
the cheapest reasonably strong CFI defenses, has a runtime overhead of
at least 15\% on the throughput of a moderately performant web server
(Apache)~\cite[Section 5.3]{cpi}. In contrast, {\erim} does not rely
on CFI for data encapsulation and has much lower overhead. Concretely,
we show in Section~\ref{erim:sec:eval} that \erim's overhead on the
throughput of a much more performant web server (NGINX) is no more
than 5\%.

The Intel MPX ISA extension~\cite{Intel2016} provides architectural
support for bounds checking needed by SFI. A compiler can use up to
four bounds registers, and each register can store a pair of 64-bit
starting and ending addresses.  Specialized instructions check a given
address and raise an exception if the bounds are violated. However,
even with MPX support, the overhead of bounds checks is of the order
of tens of percent points in many applications
(Section~\ref{sec:eval:comparison}
and~\cite{Carr:2017:DataShield,Koning2017,Oleksenko2017}).

%\todo{Reviewer A: additional citation
%DataShield - here both Memsentry and DataShield were added, since
%both demonstrate tens of percent overhead}

%% By itself, MPX cannot enforce security invariants, but it can aid
%% SFI-based techniques for data encapsulation~\cite{Koning2017}.

\paragraph{Hardware-based trusted execution environments}
Intel's SGX~\cite{INTELSGX} and ARM's TrustZone~\cite{ARMTZ} ISA
extensions allow (components of) applications to execute with
hardware-enforced isolation. JITGuard~\cite{frassetto2017jitguard},
for instance, uses SGX to protect the internal data structures of a
just-in-time compiler from untrusted code, thus preventing
code-injection attacks. While SGX and TrustZone can isolate data even
from the operating system, switching overheads are similar to other
hardware-based isolation mechanisms~\cite{Koning2017}.

IMIX~\cite{IMIX} and MicroStach~\cite{mogosanu2018microstache} propose
minimal extensions to the x86 ISA, adding load and store instructions
to access secrets in a safe region. The extended ISA can provide data
encapsulation. Both systems provide compilers that automatically
partition secrets. However, for data encapsulation in the face of
control-flow hijack attacks, both systems require CFI. As mentioned,
CFI techniques have nontrivial overhead. \erim, on the other hand,
provides strong isolation without relying on CFI and has lower
overhead.

\paragraph{ASLR}
Address space layout randomization (ASLR) is widely used to mitigate
code-reuse exploits such as those based on buffer overflow
attacks~\cite{Shacham2004, Homescu2013}.
%% By randomizing the layout of
%% code in an address space, ASLR makes it difficult for attackers to
%% reuse code as part of an exploit.
ASLR has also been used for data
encapsulation by randomizing data layout. For example, as one of the
isolation techniques used in CPI~\cite{cpi,SafeStack}, a region of
sensitive data is allocated at a random address within the 48-bit
x86-64 address space and its base address is stored in a segment
descriptor. All pointers stored in memory are offsets into the region
and do not reveal its actual address. However, all forms of ASLR are
vulnerable to attacks like thread
spraying~\cite{Shacham2004,Hund2013,Evans2015,Goktacs2016,Oikonomopoulos2016}.
Consequently, ASLR is not viable for strong memory isolation, despite
proposals such as~\cite{Lu2015} to harden it.

%% ~\cite{Evans2015,Goktacs2016,Oikonomopoulos2016}.

%% In practice, ASLR is vulnerable to attacks that
%% learn the location of code and data
%% dynamically

\paragraph{ARM memory domains}
ARM memory domains~\cite{ARM-domains} are similar to Intel MPK, the
x86 feature that {\erim} relies on. However, unlike in MPK,
changing domains is a kernel operation in ARM. Therefore,
unlike MPK, ARM's memory domains do not support low-cost user-mode
switching.

\paragraph{MPK-based techniques}
%% \updated{We start with three closely related isolation
%%   techniques~\cite{Koning2017, janus, shininglightonshadowstacks} that
%%   also rely on Intel MPK for protection.}

Koning et al.~\cite{Koning2017} present MemSentry, a general framework
for data encapsulation, implemented as a pass in the LLVM compiler
toolchain. They instantiate the framework with several different
memory isolation techniques, including many described above and one
based on MPK domains. However, MemSentry's MPK instance is secure only
with a separate defense against control-flow hijack/code-reuse attacks
to prevent adversarial misuse of PKRU-updating instructions in the
binary. Such defenses have significant overhead of their own. As a
result, the overall overhead of MemSentry's MPK instance is
significantly higher than that of \erim, which does not rely on a
defense against control-flow hijacks.
  
  %% survey techniques for efficient data
  %% encapsulation within a process, including SFI, dynamic encryption of
  %% private data using the Intel AES-NI ISA extensions, approaches that
  %% use VT-x virtualization hardware, and those that rely on the Intel
  %% MPX and MPK ISA extensions. They present a general isolation
  %% technique, called MemSentry, which instruments programs using an
  %% LLVM pass. MemSentry relies on (and assumes the existence of) a
  %% general defense against control flow hijacks to prevent untrusted
  %% code from exploiting the WRPKRU instruction to raise its
  %% privileges. However, preventing control flow hijacks adds
  %% significant runtime overhead to check indirect jumps, indirect calls
  %% and returns. For example, code-pointer integrity or CPI, one of the
  %% cheapest reasonably complete CFI approaches, has a runtime overhead
  %% of at least 15\% on even a moderate throughput web server
  %% (Apache)~\cite[Section 5.3]{cpi}. In systems such as MemSentry,
  %% where control-flow integrity is a desirable goal for
  %% \emph{independent} reasons, this additional overhead is obviously
  %% acceptable. For other applications, where control-flow integrity is
  %% not a basic goal, this overhead is a substantial burden. In
  %% contrast, {\erim} provides memory isolation without relying on
  %% control-flow integrity. We show in Section~\ref{erim:sec:eval} that
  %% {\erim} can isolate even session keys in nginx, a web server that
  %% has higher throughput than Apache, with an overhead of no more than
  %% 5\%. \todo{add hint to emulating WRPKRU, as memsentry overestimated
  %%   the cost quite a bit.}

In concurrent work~\cite{janusATC19}, Hedayati \emph{et al.}\ describe
how to isolate userspace libraries using VMFUNC or Intel MPK. The
MPK-based method is similar to {\erim}, but does not address the
challenge of ensuring that there are no exploitable occurrences of
PKRU-modifying instructions. Rewriting binaries in this manner is a
key contribution of our work (Section~\ref{sec:rewriting}). Finally,
Hedayati \emph{et al.}\ rely on kernel changes while {\erim} can run
safely on a stock Linux kernel.

%\todo{Reviewer C: compare to related work libmpk}
libmpk~\cite{park2018libmpk} virtualizes MPK memory domains
  beyond the 16 supported in hardware. It also addresses potential
  security issues in the API of Linux's MPK support. libmpk addresses
  concerns orthogonal to {\erim} because neither limitation is
  relevant to {\erim}'s use of MPK. libmpk could be combined with
  {\erim} in applications that require more than 16 components, but
  the integration remains as future work.
%  While libmpk assumes an adversary without control-flow hijack
%  capabilities, {\erim} protects against a general userspace
  % adversary.  
  
  %% concurrently to us suggest to isolate distrusting data-plane
  %% libraries using Intel MPK or VMFunc. Their technique, similar to
  %% \erim, consists of an executable memory scan before the application
  %% start and call gates to switch between memory domains. In contrast
  %% to \erim, their system Janus misses a key contribution of
  %% \erim\ which allows to rewrite binary code to safely use WRPKRU
  %% instructions. Janus relies on the absence of WRPKRU or VMFUNC
  %% instructions outside of the call gate disallowing applications to
  %% use these CPU features. Janus demonstrates how to extend the call
  %% gate mechanism to multiple distrusting domains without showing a use
  %% case that could not also be accomplished by \erim's single trusted
  %% domain. Janus also discusses Intel MPK's potential ability to defend
  %% against side-channel and microarchitectural attacks. We demonstrate
  %% Intel MPK's capabilities by showing two proof-of-concepts in
  %% section~\ref{sec:threatmodel}. \todo{maybe talk about that they have
  %%   multiple distrusting domains, requiring different stacks - we
  %%   could play the same trick with \erim}}

In recent work, Burow \emph{et al.}~\cite{shininglightonshadowstacks}
survey implementation techniques for shadow stacks. In particular,
they examine the use of MPK for protecting the integrity of shadow
stacks. Burow \emph{et al.}'s measurements of MPK overheads (Fig.\ 10
in~\cite{shininglightonshadowstacks}) are consistent with ours. Their
use of MPK could be a specific use-case for {\erim}, which is a more
general framework for memory isolation.

\section{Design}
\label{sec:design}

\paragraph{Goals}
{\erim} enables efficient data isolation within a user-space
process. Like prior work, it enables a (trusted) application component
to isolate its sensitive data from untrusted components. Unlike prior
work, {\erim} supports such isolation with \emph{low overhead} even at
\emph{high switching rates} between components \emph{without requiring
  control-flow integrity}.
In the following, we focus on the case of two components that are
isolated from each other within a single-threaded process. Later, we
describe generalizations to multi-threaded processes, more than two
components per process, and read-only sharing among components.

We use the letter {\tlib} to denote a trusted component and {\uapp} to
denote the remaining, untrusted application component.  {\erim}'s key
primitive is memory isolation: it reserves a region of the address
space and makes it accessible exclusively from the trusted component
{\tlib}. This reserved region is denoted {\tmem} and can be used by
{\tlib} to store sensitive data. The rest of the address space,
denoted {\umem}, holds the application's regular heap and stack and is
accessible from both {\uapp} and {\tlib}.  {\erim} enforces the
following invariants:\\
(1) While control is in {\uapp}, access to {\tmem} remains disabled.\\
(2) Access to {\tmem} is enabled atomically with a control transfer to
a designated entry point in {\tlib} and disabled when {\tlib}
transfers control back to {\uapp}.\\
The first invariant provides isolation of {\tmem} from {\uapp}, while
the second invariant prevents {\uapp} from confusing {\tlib} into
accessing {\tmem} improperly by jumping into the middle of {\tmem}'s
code.

\paragraph*{Background: Intel MPK}
To realize its goals, {\erim} uses the recent MPK extension to the x86
ISA~\cite{Intel2016}. With MPK, each virtual page of a process can be
associated with one of 16 protection keys, thus partitioning the
address space into up to 16 \emph{domains}. A new register, PKRU, that
is local to each logical core, determines the current access
permissions (read, write, neither or both) on each domain for the code
running on that core.
%\todo{Reviewer A: explain what the constants are in the call gates}
%\updated{To grant read or write access to domain $i$, bit $2i$ and
%  $2i+1$ have to be enabled, respectively. }
Access checks against the PKRU are implemented in hardware and
impose no overhead on program execution.

Changing access privileges requires writing new permissions to the
PKRU register with a \emph{user-mode} instruction, WRPKRU. This
instruction is relatively fast (11--260 cycles on current Intel CPUs),
does not require a syscall, changes to page tables, a TLB flush, or
inter-core synchronization.

The PKRU register can also be modified by the XRSTOR
  instruction by setting a specific bit in the eax register prior to
  the instruction (XRSTOR is used to restore the CPU's
  previously-saved extended state during a context switch).

For strong security, {\erim} must ensure that untrusted code cannot
exploit WRPKRU or XRSTOR instructions in executable pages to elevate
privileges. To this end, {\erim} combines MPK with binary inspection
to ensure that all executable occurrences of WRPKRU or XRSTOR are
\emph{safe}, i.e., they cannot be exploited to improperly elevate
privilege.

\paragraph{Background: Linux support for MPK}
As of version 4.6, the mainstream Linux kernel supports
MPK. Page-table entries are tagged with MPK domains, there are
additional syscall options to associate pages with specific domains,
and the PKRU register is saved and restored during context
switches. Since hardware PKRU checks are disabled in kernel mode, the
kernel checks PKRU permissions explicitly before dereferencing any
userspace pointer. To avoid executing a signal handler with
inappropriate privileges, the kernel updates the PKRU register to its
initial set of privileges (access only to domain 0) before
delivering a signal to a process.

%%%%%%%%%%%%%%%%%%%%%%%%%%%%%%%%%%%%%%%%%%%%%%%%%%%%%%%%%

%%%%%%%%%%%%%%%%%%%%%%%%%%%%%%%%%%%%%%%%%%%%%%%%%%%%%%%%%

\subsection{High-level design overview}
\label{sec:design:overview}

{\erim} can be configured to provide either complete isolation of
{\tmem} from {\uapp} (confidentiality and integrity), or only write
protection (only integrity). We describe the design for complete
isolation first. Section~\ref{sec:design:integrity-only} explains a
slight design re-configuration that provides only write protection.

{\erim}'s isolation mechanism is conceptually simple: It maps
{\tlib}'s reserved memory, {\tmem}, and the application's general
memory, {\umem}, to two different MPK domains. It manages MPK
permissions (the PKRU registers) to ensure that {\umem} is always
accessible, while only {\umem} is accessible when control is in
{\uapp}. It allows {\uapp} to securely transfer control to {\tlib} and
back via \emph{call gates}. A call gate enables access to {\tmem}
using the WRPKRU instruction and immediately transfers control to a
specified entry point of {\tlib}, which may be an explicit or inlined
function. When {\tlib} is done executing, the call gate disables
access to {\tmem} and returns control to {\uapp}. This enforces
{\erim}'s two invariants (1) and (2) from
Section~\ref{sec:design}. Call gates operate entirely in user-mode
(they don't use syscalls) and are described in
Section~\ref{sec:call-gate}.

\paragraph{Preventing exploitation}
A key difficulty in {\erim}'s design is preventing the untrusted
{\uapp} from exploiting occurrences of the WRPKRU or XRSTOR
instruction sequence on executable pages to elevate its
privileges. For instance, if the sequence appeared at any byte address
on an executable page, it could be exploited using control-flow hijack
attacks. To prevent such exploits, {\erim} relies on \emph{binary
  inspection} to enforce the invariant that only \emph{safe} WRPKRU
and XRSTOR occurrences appear on executable pages.

A WRPKRU occurrence is safe if it is immediately followed by one of
the following: (A) a pre-designated entry point of {\tlib}, or (B) a
specific sequence of instructions that checks that the permissions set
by WRPKRU do not include access to {\tmem} and terminates the program
otherwise. A safe WRPKRU occurrence cannot be exploited to access
{\tmem} inappropriately. If the occurrence satisfies (A), then it does
not give control to {\uapp} at all; instead, it enters {\tlib} at a
designated entry point. If the occurrence satisfies (B), then it would
terminate the program immediately when exploited to enable access to
{\tmem}.

A XRSTOR is safe if it is immediately followed by a specific
  sequence of instructions to check that the eax bit that causes
  XRSTOR to load the PKRU register is not set. Such a XRSTOR cannot be
  used to change privilege and continue execution.\footnote{We know of
    only one user-mode Linux application -- the dynamic linker, {\tt
      ld}, that legitimately uses XRSTOR. However, {\tt ld}
    categorically does \emph{not} restore PKRU through XRSTOR, so this
    safe check can be added to it.}

{\erim}'s call gates use only safe WRPKRU occurrences (and do not use
XRSTOR at all). So, they pass the binary
inspection. Section~\ref{sec:binary-inspection} describes {\erim}'s
binary inspection.

\paragraph{Creating safe binaries}
An important question is how to construct binaries that do not have
unsafe WRPKRUs and XRSTORs. On x86, these instructions may arise
inadvertently spanning the bytes of adjacent instructions or as a
subsequence in a longer instruction. To eliminate such inadvertent
occurrences, we describe a binary rewriting mechanism that rewrites
any sequence of instructions containing a WRPKRU or XRSTOR to a
functionally equivalent sequence without any WRPKRUs and XRSTORs. The
mechanism can be deployed as a compiler pass or integrated with our
binary inspection, as explained in Section~\ref{sec:rewriting}.

\subsection{Threat model}
\label{sec:threatmodel}
{\erim} makes no assumptions about the untrusted component ({\uapp})
of an application. {\uapp} may behave arbitrarily and may contain
memory corruption and control-flow hijack vulnerabilities that may be
exploited during its execution.
%, and it may attempt to map new executable code.

However, {\erim} assumes that the trusted component {\tlib}'s binary
does not have such vulnerabilities and does not compromise sensitive
data through explicit information leaks, by calling back into {\uapp}
while access to {\tmem} is enabled, or by mapping executable pages
with unsafe/exploitable occurrences of the WRPKRU or XRSTOR
instruction.
%, WRPKRU (we define safe occurrences of this instruction in
%Section~\ref{sec:design:overview}).

The hardware, the OS kernel, and a small library added by {\erim} to
each process that uses {\erim} are trusted to be secure. We also
assume that the kernel enforces standard DEP---an executable page must
not be simultaneously mapped with write permissions. {\erim} relies on
a list of legitimate entry points into {\tlib} provided either by the
programmer or the compiler, and this list is assumed to be correct
(see Section~\ref{sec:binary-inspection}).  The OS's dynamic program
loader/linker is trusted to invoke {\erim}'s initialization function
before any other code in a new process.

Side-channel and rowhammer attacks, and microachitectural leaks,
although important, are beyond the scope of this work. However,
\erim\ is compatible with existing defenses. Our current
  \emph{prototype} of {\erim} is incompatible with applications that
  simultaneously use MPK for other purposes, but this is not
  fundamental to {\erim}'s design. Such incompatibilities can be
  resolved as long as the application does not re-use the MPK domain
  that {\erim} reserves for {\tlib}.

\subsection{Call gates}
\label{sec:call-gate}

A call gate transfers control from {\uapp} to {\tlib} by enabling
access to {\tmem} and executing from a designated entry point of
{\tlib}, and later returns control to {\uapp} after disabling access
to {\tmem}. This requires two WRPKRUs. The primary challenge in
designing the call gate is ensuring that both these WRPKRUs are safe
in the sense explained in Section~\ref{sec:design:overview}.

%% Next, we describe {\erim}'s call gate implementation. Recall that when
%% {\uapp} transfers control to {\tlib}, the call gate executes WRPKRU to
%% enable access to {\tmem} and then transfers control to {\tlib}'s
%% pre-designated entry point. When {\tlib} is done processing, the call
%% gate disables access to {\tmem} and transfers control back to
%% {\uapp}. The call gate has to be robust to control flow hijacks:
%% {\uapp} should not be able to exploit the WRPKRU occurrences in the
%% call gate to obtain access to {\tmem}.

%% \begin{lstlisting} [float=*,caption={Call gate implementation in pseudocode. The code of the trusted component's entry point may be inlined by the compiler on line 3, or there may be an explicit call to it.},label={lst:call-gate},frame=rltb,numbers=left,numberstyle=\small,numberblanklines=false,stepnumber=1,showlines=true]
%% eax <- ACCESS_TRUSTED | ACCESS_UNTRUSTED
%% WRPKRU // copies eax to PKRU
%% ... execute starting from the trusted component's entry point here ...
%% eax <- ACCESS_UNTRUSTED // executes after trusted component is done
%% WRPKRU
%% if (eax & ACCESS_TRUSTED) then terminate() // prevents exploitation of WRPKRU
%% \end{lstlisting}

\begin{lstlisting}[basicstyle=\small\ttfamily,float=t,caption={Call gate in assembly. The code of the trusted component's entry point may be inlined by the compiler on line 6, or there may be an explicit direct call to it.},label={lst:call-gate},frame=tb,numbers=right,numberstyle=\small,numberblanklines=false,stepnumber=1,showlines=true,numbersep=-10pt]
  xor ecx, ecx
  xor edx, edx
  mov PKRU_ALLOW_TRUSTED, eax
  WRPKRU // copies eax to PKRU

// Execute trusted component's code 

  xor ecx, ecx
  xor edx, edx
  mov PKRU_DISALLOW_TRUSTED, eax
  WRPKRU // copies eax to PKRU
  cmp PKRU_DISALLOW_TRUSTED, eax
  je continue
  syscall exit // terminate program
continue:
// control returns to the untrusted application here  
\end{lstlisting}

Listing~\ref{lst:call-gate} shows the assembly code of a call gate.
WRPKRU expects the new PKRU value in the \lstinline{eax} register and
requires ecx and edx to be $0$. The call gate works as follows. First,
it sets PKRU to enable access to {\tmem} (lines 1--4). 
%\todo{Reviewer A: explain what the constant values mean}
The macro PKRU\_ALLOW\_TRUSTED is a constant that allows access to
{\tmem} and {\umem}.\footnote{To grant read (resp.\ write) access to
  domain $i$, bit $2i$ (resp.\ $2i+1$) must be set in the
  PKRU. PKRU\_ALLOW\_TRUSTED sets the 4 least significant bits to
  grant read and write access to domains 0 ({\umem}) and 1 ({\tmem}).}
Next, the call gate transfers control to the designated entry point of
{\tlib} (line 6).  {\tlib}'s code may be invoked either by a direct
call, or it may be inlined.

After {\tlib} has finished, the call gate sets PKRU to disable access
to {\tmem} (lines 8--11). The macro PKRU\_DISALLOW\_TRUSTED is a
constant that allows access to {\umem} but not {\tmem}.  Next, the
call gate checks that the PKRU was actually loaded with
PKRU\_DISALLOW\_TRUSTED (line 12). If this is not the case, it
terminates the program (line 14), else it returns control to {\uapp}
(lines 15--16). The check on line~12 may seem redundant since
\lstinline{eax} is set to PKRU\_DISALLOW\_TRUSTED on line 10. However,
the check prevents \emph{exploitation} of the WRPKRU on line~11 by a
control-flow hijack attack (explained next).

\paragraph*{Safety}
Both occurrences of WRPKRU in the call gate are safe. Neither can be
exploited by a control flow hijack to get unauthorized access to
{\tmem}. The first occurrence of WRPKRU (line~4)
%is of form (A) and
is immediately followed by (a direct control transfer to) a designated
entry point of {\tlib}. This instance cannot be exploited to transfer
control to anywhere else. The second occurrence of WRPKRU (line~11) is
followed by a check that terminates the program if the new permissions
include access to {\tmem}. If, as part of an attack, the execution
jumped directly to line 11 with any value other than
PKRU\_DISALLOW\_TRUSTED in eax, the program would be terminated on
line 14.

%% In fact, lines 12--15 constitute the
%% \emph{exact} suffix that must follow a WRPKRU for the WRPKRU to be
%% classified safe under condition (B) in our binary inspection
%% mechanism.

%% An obvious security concern is that a control flow hijack may try to
%% exploit the two occurrences of WRPKRU in {\erim}'s call gate by
%% setting \lstinline{eax} to a permission that allows access to
%%   {\tmem}, and then immediately jumping to one of the two occurrences
%%   of WRPKRU.
%% %
%% However, this kind of attack affords no advantage to the
%% adversary. The occurrence of WRPKRU on line~2 is immediately followed
%% by a control transfer to {\tlib}'s designated entry point and, hence,
%% does not allow the adversary to execute anything except {\tlib} with
%% elevated privilege. The occurrence of WRPKRU on line~5 is immediately
%% followed by a check that terminates the program if \lstinline{eax}
%%   contains access to {\tmem}. Consequently, this occurrence of WRPKRU
%%   in the call gate also cannot be exploited by an adversary to execute
%%   code with elevated privilege.

\paragraph*{Efficiency}
A call gate's overhead on a roundtrip from {\uapp} to {\tlib} is two
WRPKRUs, a few very fast, standard register operations and one
conditional branch instruction.  This overhead is very low compared to
other hardware isolation techniques that rely on pages tables and
syscalls or hypervisor trampolines to change privileges (see also
Section~\ref{sec:eval:comparison}).

\paragraph*{Use considerations}
{\erim}'s call gate omits features that readers may expect. These
features have been omitted to avoid having to pay their overhead when
they are not needed.
First, the call gate does not include support to pass parameters from
{\uapp} to {\tlib} or to pass a result from {\tlib} to {\uapp}. These
can be passed via a designated shared buffer in {\umem} (both
{\uapp} and {\tlib} have access to {\umem}).
Second, the call gate does not scrub registers when switching from
{\tlib} to {\uapp}. So, if {\tlib} uses confidential data, it should
scrub any secrets from registers before returning to {\uapp}.
Further, because {\tlib} and {\uapp} share the call stack, {\tlib}
must also scrub secrets from the stack prior to
returning. Alternatively, {\tlib} can allocate a private stack for
itself in {\tmem}, and {\tlib}'s entry point can switch to that stack
immediately upon entry. This prevents {\tlib}'s secrets from being
written to {\uapp}'s stack in the first place. (A private stack is
also necessary for multi-threaded applications; see
Section~\ref{sec:design:multi-threads}).
%% Call gates with any subset of these features can be generated by an
%% {\erim}-aware compiler or provided as part of the {\erim} library.

%%%%%%%%%%%%%%%%%%%%%%%%%%%%%%%%%%%%%%%%%%%%%%%%%%%%%%%%%%%%%%

\subsection{Binary inspection}
\label{sec:binary-inspection}

Next, we describe {\erim}'s binary inspection. The inspection prevents
{\uapp} from mapping any executable pages with unsafe WRPKRU and
XRSTOR occurrences and consists of two parts: (i) an inspection
function that verifies that a sequence of pages does not contain
unsafe occurrences; and, (ii) an interception mechanism that prevents
{\uapp} from mapping executable pages without inspection.

\paragraph{Inspection function}
The inspection function \emph{scans} a sequence of pages for
instances of WRPKRU and XRSTOR. It also inspects any adjacent
executable pages in the address space for instances that cross a
page boundary.

For every WRPKRU, it checks that the WRPKRU is safe, i.e., either
condition (A) or (B) from Section~\ref{sec:design:overview}
holds.
To check for condition (A), {\erim} needs a list of designated entry
points of {\tlib}. The source of this list depends on the nature of
{\tlib} and is trusted. If {\tlib} consists of library functions, then
the programmer marks these functions, e.g., by including a unique
character sequence in their names. If the functions are not inlined by
the compiler, their names will appear in the symbol table. If
{\tlib}'s functions are subject to inlining or if they are generated
by a compiler pass, then the compiler must be directed to add their
entry locations to the symbol table with the unique character
sequence. In all cases, {\erim} can identify designated entry points
by looking at the symbol table and make them available to the
inspection function.

Condition (B) is checked easily by verifying that the WRPKRU is
immediately followed by \emph{exactly} the instructions on lines
12--15 of Listing~\ref{lst:call-gate}. These instructions ensure that
the WRPKRU cannot be used to enable access to {\tmem} and continue
execution.

For every XRSTOR, the inspection function checks that the XRSTOR is
followed immediately by the following instructions, which check that
the eax bit that causes XRSTOR to load PKRU (bit 9) is not set:
\texttt{bt eax, 0x9; jnc .safe; EXIT; .safe:\ldots}. Here,
\texttt{EXIT} is a macro that exits the program. Trivially, such a
XRSTOR cannot be used to enable access to {\tmem} and continue
execution.

\paragraph{Interception}
On recent ($\geq$\,4.6) versions of Linux, interception can be
implemented \emph{without kernel changes}. We install a seccomp-bpf
filter~\cite{seccompbpf} that catches mmap, mprotect, and
pkey\_mprotect syscalls which attempt to map a region of memory as
executable (mode argument {\tt PROT\_EXEC}). Since the bpf filtering
language currently has no provisions for reading the PKRU register, we
rely on seccomp-bpf's {\tt SECCOMP\_RET\_TRACE} option to notify a
ptrace()-based tracer process. The tracer inspects the tracee and
allows the syscall if it was invoked from {\tlib} and denies it
otherwise. The tracer process is configured so that it traces any child
of the tracee process as well. While ptrace() interception is
expensive, note that it is required only when a program maps pages as
executable, which is normally an infrequent operation.

If programs map executable pages frequently, a more efficient
interception can be implemented with a simple Linux Security Module
(LSM)~\cite{lsm}, which allows mmap, mprotect and pkey\_mprotect
system calls only from {\tlib}.  (Whether such a call is made by
{\uapp} or {\tlib} is easily determined by examining the PKRU register
value at the time of the syscall.) Our prototype uses this
implementation of interception.  Another approach is to implement a
small (8 LoC) change to seccomp-bpf in the Linux kernel, which allows
a bpf filter to inspect the value of the PKRU register. With this
change in place, we can install a bpf filter that allows certain
syscalls only from {\tlib}, similar to the LSM module.

With either interception approach in place, {\uapp} must go through
{\tlib} to map executable pages. {\tlib} maps the pages only after
they have passed the inspection function.
%
%We implemented a third approach, which requires a slightly larger
%(\todo{103 LoC) change to the Linux kernel, but has the advantage that
%executable pages can be scanned on demand when they are first
%executed.
Regardless of the interception method, pages can be inspected upfront
when {\tlib} attempts to map them as executable, or on demand when
they are executed for the first time.

On-demand inspection is preferable when a program maps a large
executable segment but eventually executes only a small number of
pages. With on-demand inspection, when the process maps a region as
executable, {\tlib} instead maps the region read-only but records that
the pages
%in the region are \emph{supposed} to be executable
are pending inspection.
When control transfers to such a page, a fault occurs. The fault traps
to a dedicated signal handler, which {\erim} installs when it
initializes (the LSM or the tracer prevents {\uapp} from overriding
this signal handler). This signal handler calls a {\tlib} function
that checks whether the faulting page is pending inspection and, if
so, inspects the page.  If the inspection passes, then the handler
remaps the page with the execute permission and resumes execution of
the faulting instruction, which will now succeed. If not, the program
is terminated.

The interception and binary inspection has very low overhead in
practice because it scans an executable page at most once.
%---when the page is first used.
It is also fully transparent to {\uapp}'s code if all WRPKRUs
and XRSTORs in the binary are already safe.

\paragraph{Security}
We briefly summarize how {\erim} attains security. The binary
inspection mechanism prevents {\uapp} from mapping any executable page
with an unsafe WRPKRU or XRSTOR. {\tlib} does not contain any
executable unsafe WRPKRU or XRSTOR by assumption. Consequently, only
safe WRPKRUs and XRSTORs are executable in the entire
address space at any point.
%% and they transfer control to one of {\tlib}'s designated entry
%% points, which are safe by assumption.
Safe WRPKRUs and XRSTORs preserve {\erim}'s two security invariants
(1) and (2) by design. Thus {\tmem} is accessible only while {\tlib}
executes starting from legitimate {\tlib} entry points.
%Hence, {\tmem} remains isolated from {\uapp}.
%% which are safe
%% by assumption. 

%%%%%%%%%%%%%%%%%%%%%%%%%%%%%%%%%%%%%%%%%%%%%%%%%%%%%%%%%%%%%%

\subsection{Lifecycle of an {\erim} process}
\label{sec:life-cycle}

%\todo{Reviewer B-D: better describe different options for deployment, 
%separate out the general flow of control vs. the actual techniques
%to deploy the flow. 1) lifecycle - independent of LD\_PRELOAD, linking,
%or compilation 2) Explain How to startup ERIM before APP 3) Instrument
%secure component by source code switches (shown in listing) or dynamic 
%linking}

%\todo{Reviewer C: Binary only deployment?}

As part of a process's initialization, before control is transferred
to main(), {\erim} creates a second MPK memory domain for {\tmem} in
addition to the process's default MPK domain, which is used for
{\umem}. {\erim} maps a memory pool for a dynamic memory allocator to
be used in {\tmem} and hooks dynamic memory allocation functions so
that invocations are transparently redirected to the appropriate pool
based on the value of the PKRU register. This redirection provides
programmer convenience but is not required for security.  If {\uapp}
were to call {\tlib}'s allocator, it would be unable to access
{\tmem}'s memory pool and generate a page fault.  Next, {\erim} scans
{\umem}'s executable memory for unsafe WRPKRUs and XRSTORs, and
installs one of the interception mechanisms described in
Section~\ref{sec:binary-inspection}.  Finally, depending on whether
main() is in {\uapp} or {\tlib}, {\erim} initializes the PKRU register
appropriately and transfers control to main(). After main() has
control, the program executes as usual. It can map, unmap and
  access data memory in {\umem} freely. However, to access {\tmem}, it
  must invoke a call gate.

\subsection{Developing {\erim} applications}
\label{sec:writeprog}

\begin{lstlisting}[basicstyle=\small\ttfamily,float=t,caption={C component isolated with \erim},label={lst:cprog},frame=tb,numbers=right,numberstyle=\small,numberblanklines=false,stepnumber=1,showlines=true,numbersep=-10pt]
typedef struct secret {
  int number; } secret;
secret* initSecret() { 
  ERIM_SWITCH_T; 
  secret * s = malloc(sizeof(secret));
  s->number = random();
  ERIM_SWITCH_U;
  return s;
}
int compute(secret* s, int m) { 
  int ret = 0;
  ERIM_SWITCH_T;
  ret = f(s->number, m);
  ERIM_SWITCH_U;
  return ret;
}
\end{lstlisting}

We describe here three methods of developing applications or modifying
existing applications to use {\erim}.

The {\em binary-only} approach requires that either {\uapp} or {\tlib}
consist of a set of functions in a dynamic link library. In this case,
the library and the remaining program can be used in unmodified binary
form. An additional {\erim} dynamic wrapper library is added using
LD\_PRELOAD, which wraps the entry points with stub functions that
implement the call gates and have names that indicate to the {\erim}
runtime the valid entry points. We have used this approach to isolate
SQLite within the Node.js runtime (Section~\ref{sec:usecases}).

The {\em source} approach requires that either {\uapp} or {\tlib}
consist of a set of functions that are not necessarily in a separate
compilation unit or library. In this case, the source code is modified
to wrap these functions with stubs that implement the call gates, and
choose names that indicate valid entry points. We used this
approach to isolate the crypto functions and session keys in OpenSSL
(Section~\ref{sec:usecases}).

The {\em compiler} approach requires modifications to the compiler to
insert call gates at appropriate points in the executable and generate
appropriate symbols that indicate valid entry points. This approach is
the most flexible because it allows arbitrary inlining of {\uapp} and
{\tlib} code. We used this approach to isolate the metadata in CPI
(Section~\ref{sec:usecases}).  

\if 0

To develop an application that uses {\erim} to isolate sensitive
memory, a developer has to complete three main tasks. First, decide
the split and interface between {\tlib} and {\uapp}. Second, swap
function calls to {\tlib} against the appropriate call gate. Third,
ensure {\erim} initializes before the application start.

In case {\tlib} are small code blocks inlined in {\uapp} or
functions within {\uapp}'s code base, 
call gates need to be inlined into the source code of {\uapp} 
automatically by a compiler or the developer. In case {\tlib}
comprises a shared library, its global functions
can be relinked at runtime to wrappers that invoke a call gate.
While inlining call gates allows to isolate small parts of code,
relinking wrappers at runtime allows to deploy {\erim} without
recompilation of {\uapp}.

To enforce {\erim}'s initialization before the application start,
we can either rely on source code changes calling {\erim} before main(), or a shared library
supplied by {\erim} who's initialization function is called before
{\uapp}'s main. Shared libraries may be added to the execution of 
any application using LD\_PRELAOD.
\fi

Next, we give a simple example describing the process of developing a
new C application using the {\em source} approach. {\erim} provides a
C library and header files to insert call gates, initialize {\erim},
and support dynamic memory allocation.
Listing~\ref{lst:cprog} demonstrates an example C program that
isolates a data structure called {\tt secret} (lines 1--2). The
structure contains an integer value. Two functions, {\tt initSecret}
and {\tt compute}, access secrets and bracket their respective
accesses with call gates using the macros {\tt ERIM\_SWITCH\_T} and
{\tt ERIM\_SWITCH\_U}.  \erim\ isolates {\tt secret} such that only
code that appears between {\tt ERIM\_SWITCH\_T} and {\tt
  ERIM\_SWITCH\_U}, i.e., code in {\tlib}, may access {\tt
  secret}. {\tt initSecret} allocates an instance of {\tt secret}
while executing inside {\tlib} by first allocating memory in {\tmem}
and then initializing the secret value. {\tt compute} computes a
function {\tt f} of the secret inside {\tlib}.

\subsection{Extensions}
\label{sec:design:integrity-only}
\label{sec:design:multi-threads}

Next, we discuss extensions to {\erim}'s basic design.

\paragraph{Multi-threaded processes} {\erim}'s basic design works as-is
with multi-threaded applications. Threads are created as usual,
e.g.\ using libpthread. The PKRU register is saved and restored by the
kernel during context switches.  However, multi-threading imposes an
additional requirement on {\tlib} (not on {\erim}): In a
multi-threaded application, it is essential that {\tlib} allocate a
private stack in {\tmem} (not {\umem}) for each thread and execute its
code on these stacks. This is easy to implement by switching stacks at
{\tlib}'s entry points. Not doing so and executing {\tlib} on standard
stacks in {\umem} runs the risk that, while a thread is executing in
{\tlib}, another thread executing in {\uapp} may corrupt or read the
first thread's stack frames. This can potentially destroy {\tlib}'s
integrity, leak its secrets and hijack control while access to {\tmem}
is enabled. By executing {\tlib}'s code on stacks in {\tmem}, such
attacks are prevented.

%%%%%%%%%%%%%%%%%%%%%%%%%%%%%%%%%%%%%%%%%%%%%%%

\paragraph{More than two components per process}

Our description of {\erim} so far has been limited to two components
({\tlib} and {\uapp}) per process. However, {\erim} generalizes easily
to support as many components as the number of domains Linux's MPK
support can provide (this could be less than 16 because the kernel may
reserve a few domains for specific purposes). Components can have
arbitrary pairwise trust relations with each other, as long as the
trust relations are transitive. A simple setting could have a default
domain that trusts all other domains (analogous to {\uapp}) and any
number of additional domains that do not trust any others. {\erim}'s
initialization code creates a private heap for each component, and
{\erim}'s custom allocator allocates from the heap of the currently
executing component. Each component can also (in its own code)
allocate a per-thread stack, to protect stack-allocated sensitive data
when calling into other untrusted domains. Stacks can be mandatorily
switched by {\erim}'s call gates.

\paragraph{{\erim} for integrity only}
Some applications care only about the integrity of protected data, but
not its confidentiality. Examples include CPI, which needs to protect
only the integrity of code pointers. In such applications, efficiency
can be improved by allowing {\uapp} to \emph{read} {\tmem} directly,
thus avoiding the need to invoke a call gate for reading {\tmem}. The
{\erim} design we have described so far can be easily modified to
support this case. Only the definition of the constant
PKRU\_DISALLOW\_TRUSTED in Listing~\ref{lst:call-gate} has to change
to also allow read-only access to {\tmem}. With this change, read
access to {\tmem} is always enabled.

%%%%%%%%%%%%%%%%%%%%%%%%%%%%%%%%%%%%%%%%%%%%%%%%%%%%%%%%%%%%%%

\paragraph{Just-in-time (jit) compilers with \erim} 
%  \todo{Reviewer B: explain Jit better} 
{\erim} works with jit compilers that follow standard DEP
    and do not allow code pages that are writable and executable at
    the same time.  Such jit compilers write new executable code into
    newly allocated, non-executable pages and change these pages'
    permissions to non-writable and executable once the compilation
    finishes. {\erim}'s mprotect interception defers enabling execute
    permissions until after a binary inspection, as described in
    Section~\ref{sec:binary-inspection}.
%  At this
%  time \erim's binary inspection mechanism maps the page without
%  execute permission. When execution reaches the newly compiled code a
% segmentation fault occurs.
  When a newly compiled page is executed for the first time, {\erim}
  handles the page execute permission fault, scans the new page for
  unsafe WRPKRUs/XRSTORs and enables the execute permission
  if no unsafe occurrences exist. This mechanism is safe, but may lead
  to program crashes if the jit compiler accidentally emits an unsafe
  WRPKRU or XRSTOR. \erim-aware jit compilers can emit
  WRPKRU- and XRSTOR-free binary code by relying on the
  rewrite strategy described in Section~\ref{sec:rewriting}, and
  inserting call gates when necessary.

%An additional optimization could
%inform \erim's binary inspection mechanism at the end of the jit
%compiler's pipeline to scan the page for WRPKRUs and enable the
%execute permission. This lowers the number of segmentation faults, but
%requires jit compilers to support \erim.

%% In addition to supporting \erim, jit compilers can prevent
%% memory-corruption attacks~\cite{gawlik2018sok} from, e.g., corrupting
%% the jit compiler's state using \erim. \erim's memory isolation can
%% efficiently protect the jit compiler's state by isolating the jit
%% compiler in the trusted domain, while the application runs in the
%% untrusted domain. As a result, \erim\ prevents the untrusted
%% application from accessing the jit compiler's state preventing
%% memory-corruption attacks. Compared to existing
%% work~\cite{frassetto2017jitguard} which relies on Intel SGX to isolate
%% the compiler's state, \erim's isolation is highly efficient.

%%%%%%%%%%%%%%%%%%%%%%%%%%%%%%%%%%%%%%%%%%%%%%%%%%%%%%%%%%%%%%

\paragraph{OS privilege separation}
The design described so far provides memory isolation.  Some
applications, however, require privilege separation between {\tlib}
and {\uapp} with respect to OS resources. For instance, an application
might need to restrict the filesystem name space accessible to {\uapp}
or restrict the system calls available to {\uapp}.

{\erim} can be easily extended to support privilege separation with
respect to OS resources, using one of the techniques described in
Section~\ref{sec:binary-inspection} for intercepting systems calls
that map executable pages. In fact, intercepting and disallowing these
system calls when invoked from {\uapp} is just a special case of
privilege separation.  During process initialization, {\erim} can
instruct the kernel to restrict {\uapp}'s access rights.  After this,
the kernel refuses to grant access to restricted resources whenever
the value of the PKRU is not PKRU\_ALLOW\_TRUSTED, indicating that the
syscall does not originate from {\tlib}.  To access restricted
resources, {\uapp} must invoke {\tlib}, which can filter syscalls.

\begin{table*}[t]
\centering
%\begin{tabularx}{0.95\textwidth}{l|X|X|X}
\small
\begin{tabular}{|l|>{\raggedright}p{0.1\textwidth}|p{0.33\textwidth}|l|p{0.33\textwidth}|}
\hline
        Overlap with & Cases & Rewrite strategy & ID & Example \\ \hline
        
        Opcode & Opcode = WRPKRU/ XRSTOR & Insert safety check after instruction & 1 & \\ \hline
        Mod R/M & Mod R/M = 0x0F & Change to unused register + move command & 2 & add ecx, [ebx + 0x01EF0000] $\rightarrow$ mov eax, ebx; add ecx,  [eax + 0x01EF0000]; \\ \cline{3-5}
        & & Push/Pop used register + move command & 3 & add ecx, [ebx + 0x01EF0000] $\rightarrow$ push eax; mov eax, ebx; add ecx, [eax + 0x01EF0000]; pop eax;\\ \hline
                        %         & Mod R/M unchanged & See SIB, displacement, immediate & 4 & \\ \hline
        %% SIB & SIB = 0x0F or SIB = 0x01 & Change scale of SIB and update displacement & 4 & \\ \cline{3-5}
        %%                      &            & Multiply index register by scale, execute opcode, divide index register & 5 &\\ \cline{2-5}
        %%                  %    & SIB \hspace{0.5cm} unchanged & See displacement or immediate & 7 &\\ \hline
        Displacement & Full/Partial sequence & Change mode to use register & 4 & add eax, 0x0F01EF00 $\rightarrow$ (push ebx;) mov ebx, 0x0F010000; add ebx, 0x0000EA00; add eax, ebx; (pop ebx;) \\ \cline{2-5}
	& Jump-like instruction & Move code segment to alter constant used in address & 5 & call [rip + 0x0F01EF00] $\rightarrow$ call [rip + 0x0FA0EEFF]\\ \hline
                                      %% & Ends with prefix      & See immediate & 8 & \\ \hline
        Immediate & Full/Partial sequence & Change mode to use register & 6 & add eax, 0x0F01EF $\rightarrow$ (push ebx;) mov ebx, 0x0F01EE00; add ebx, 0x00000100; add eax, ebx; (pop ebx;) \\ \cline{2-5}
                  & Associative opcode  & Apply instruction twice with different immediates to get equivalent effect & 7 & add ebx, 0x0F01EF00 $\rightarrow$ add ebx, 0x0E01EF00; add ebx, 0x01000000\\ \hline

%\end{tabularx}
\end{tabular}
\caption{Rewrite strategy for intra-instruction occurrences of WRPKRU and XRSTOR}
\label{tab:rewritestrategies}
\end{table*}

\section{Rewriting program binaries}
\label{sec:rewriting}

The binary inspection described in Section~\ref{sec:binary-inspection}
guarantees that executable pages do not contain unsafe instances of
the WRPKRU and XRSTOR instructions. This is \emph{sufficient} for
{\erim}'s safety. In this section, we show how to generate or modify
program binaries to not contain unsafe WRPKRUs and XRSTORs, so that
they pass the binary inspection.

Intentional occurrences of WRPKRU that are not immediately followed by
a transfer to {\tlib} and all occurrences of XRSTOR, whether they are
generated by a compiler or written manually in assembly, can
be made safe by inserting the checks described in
Section~\ref{sec:binary-inspection} after the instances.  Inadvertent
occurrences---those that arise unintentionally as part of a longer x86
instruction and operand, or spanning two consecutive x86
instructions/operands---are more interesting. We describe a rewrite
strategy to eliminate such occurrences and how the strategy can be
applied by a compiler or a binary rewriting tool.  The strategy
%can be shown to be \emph{complete}: It
can rewrite any sequence of x86 instructions and operands containing
an inadvertent WRPKRU or XRSTOR to a functionally equivalent sequence
without either. In the following we describe the strategy, briefly
argue why it is complete, and summarize an empirical evaluation of its
effectiveness.
%\todo{Reviewer A: provide stronger argument, why rewriting is complete}

\paragraph*{Rewrite strategy}
WRPKRU is a 3 byte instruction, 0x0F01EF. XRSTOR is also always a
3-byte instruction, but it has more variants, fully described by the
regular expression 0x0FAE[2|6|A][8-F]. There are two cases to
consider. First, a WRPKRU or XRSTOR sequence can span two or more x86
instructions. Such sequences can be ``broken'' by inserting a 1-byte
nop like 0x90 between the two consecutive instructions. 0x90 does not
coincide with any individual byte of WRPKRU or XRSTOR, so this
insertion cannot generate a new occurrence.

Second, a WRPKRU or XRSTOR may appear entirely within a longer
instruction including any immediate operand. Such cases can be
rewritten by replacing them with a semantically equivalent instruction
or sequence of instructions. Doing so systematically requires an
understanding of x86 instruction coding. An x86 instruction contains:
(i) an opcode field possibly with prefix, (ii) a MOD R/M field that
determines the addressing mode and includes a register operand, (iii)
an optional SIB field that specifies registers for indirect memory
addressing, and (iv) optional displacement and/or immediate fields
that specify constant offsets for memory operations and other constant
operands.

The strategy for rewriting an instruction depends on the fields with
which the WRPKRU or XRSTOR subsequence
overlaps. Table~\ref{tab:rewritestrategies} shows the complete
strategy.

An opcode field is at most 3-bytes long. If the WRPKRU (XRSTOR) starts
at the first byte, the instruction \emph{is} WRPKRU (XRSTOR). In
this case, we make the instruction safe by inserting the corresponding
check from Section~\ref{sec:binary-inspection} after it. If the
WRPKRU or XRSTOR starts after the first byte of the opcode,
it must also overlap with a later field. In this case, we rewrite
according to the rule for that field below.

If the sequence overlaps with the MOD R/M field, we change the
register in the MOD R/M field. This requires a free register. If one
does not exist, we rewrite to push an existing register to the stack,
use it in the instruction, and pop it back. (See lines 2 and 3 in
Table~\ref{tab:rewritestrategies}.)

If the sequence overlaps with the displacement or the immediate field,
we change the mode of the instruction to use a register instead of a
constant. The constant is computed in the register before the
instruction (lines 4 and 6). If a free register is unavailable, we
push and pop one. Two instruction-specific optimizations are
possible. First, for jump-like instructions, the jump target can be
relocated in the binary; this changes the displacement in the
instruction, obviating the need a free register (line~5). Second,
associative operations like addition can be performed in two
increments without an extra register (line~7).  Rewriting the SIB
field is never required because any WRPKRU or XRSTOR must
overlap with at least one non-SIB field (the SIB field is 1 byte long
while these instructions are 3 bytes long).

Compilers and well-written assembly programs normally do not mix data
like constants, jump tables, etc. with the instruction stream and
instead place such data in a non-executable data segment. If so,
WRPKRU or XRSTOR sequences that occur in such data can be ignored.
%internally or at the
%boundary with instructions, it should be (and normally is) placed in a
%non-executable data segment. 

%I would think so. \todo{Confirm that the previous paragraph is still true for XRSTOR.}

%Yes this was a copy issue. \todo{Check the hex numbers in the last column of ID 5. I don't see any occurrence of WRPKRU in there. Seems to be some typo.}

%as explained in my email \todo{Check that the rewritten sequences in the last column do not contain XRSTORs. If it helps, the key rule for XRSTOR is that the 'reg' field of modR/M should not be ECX or EBP. Some of the rewritten examples do use ECX but I am not sure if it ends up in the reg field or elsewhere.}

\paragraph*{Compiler support}
For binaries that can be recompiled from source, rewriting can be
added to the codegen phase of the compiler, which converts the
intermediate representation (IR) to machine instructions. Whenever
codegen outputs an inadvertent WRPKRU or XRSTOR, the surrounding
instructions in the IR can be replaced with equivalent instructions as
described above, and codegen can be run again.

\paragraph*{Runtime binary rewriting}
For binaries that cannot be recompiled, binary rewriting can be
integrated with the interception and inspection mechanism
(Section~\ref{sec:binary-inspection}). When the inspection discovers
an unsafe WRPKRU or XRSTOR on an executable page during its
scan, it overwrites the page with 1-byte traps, makes it executable,
and stores the original page in reserve without enabling it for
execution. Later, if there is a jump into the executable page, a trap
occurs and the trap handler discovers an entry point into the page.

The rewriter then disassembles the \emph{reserved page} from that
entry point on, rewriting any discovered WRPKRU or XRSTOR occurrences,
and copies the rewritten instruction sequences back to the executable
page. To prevent other threads from executing partially overwritten
instruction sequences, we actually rewrite a fresh copy of the
executable page with the new sequences, and then swap this rewritten
copy for the executable page. This technique is transparent to the
application, has an overhead proportional to the number of entry
points in offending pages (it disassembles from every entry point only
once) and maintains the invariant that only safe pages are executable.

A rewritten instruction sequence is typically longer than the original
sequence and therefore cannot be rewritten in-place. In this
case, binary rewriting tools place the rewritten sequence on a new
page, replace the first instruction in the original sequence with a
direct jump to the rewritten sequence, and insert a direct jump back
to the instruction following the original sequence after the rewritten
sequence. Both pages are then enabled for execution.

%% \updated{
%% As an alternative to integrating binary rewriting with the
%% interception/inspection mechanism, binaries can be disassembled and
%% rewritten \emph{statically}. We implemented this to test our rewrite
%% strategy, as described next.
%% %
%% }

%\todo{Reviewer D: Concerned about corner cases when rewriting at runtime.
%Particularly when code is moved - which we never do - explain how
%rewriting works}

\paragraph*{Implementation and testing}
The rewrite strategy is arguably complete. We have implemented the
strategy
%for WRPKRU/XRSTOR
as a library, which can be used either with the inspection mechanism
as explained above or with a \emph{static} binary rewrite tool, as
described here. To gain confidence in our implementation, we examined
all binaries of five large Linux distributions (a total of 204,370
binaries). Across all binaries, we found a total of 1213 WRPKRU/XRSTOR
occurrences in code segments. We then used a standard tool,
Dyninst~\cite{dyninst}, to try to disassemble and rewrite these
occurrences. Dyninst was able to disassemble 1023 occurrences and, as
expected, our rewriter rewrote all instances successfully. Next, we
wanted to run these 1023 rewritten instances. However, this was
infeasible since we did not know what inputs to the binaries would
cause control to reach the rewritten instances. Hence, we constructed
two hand-crafted binaries with WRPKRUs/XRSTORs similar to the 1023
occurrences, rewrote those WRPKRUs/XRSTORs with Dyninst and checked
that those rewritten instances ran correctly. Based on these
experiments, we are confident that our implementation of WRPKRU/XRSTOR
rewriting is robust.

\section{Use Cases}
\label{sec:usecases}

{\erim} goes beyond prior work by providing efficient isolation with
very high component switch rates of the order of $10^5$ or $10^6$
times a second.  We describe three such use cases here, and report
{\erim}'s overhead on them in Section~\ref{erim:sec:eval}.

\paragraph*{Isolating cryptographic keys in web servers}
Isolating \emph{long-term} SSL keys to protect from web server
vulnerabilities such as the Heartbleed bug~\cite{heartbleed} is
well-studied~\cite{Litton2016,Liu2015}. However, long-term keys are
accessed relatively infrequently, typically only a few times per user
session.  \emph{Session keys}, on the other hand, are accessed far
more frequently---over $10^6$ times a second per core in a high
throughput web server like NGINX. Isolating sessions keys is relevant
because these keys protect the confidentiality of individual
users. With its low-cost switching, {\erim} can be used to isolate
session keys efficiently. To verify this, we partitioned OpenSSL's
low-level crypto library (libcrypto) to isolate the session keys and
basic crypto routines, which run as {\tlib}, from the rest of the web
server, which runs as {\uapp}.

\paragraph*{Native libraries in managed runtimes}
Managed runtimes such as a Java or JavaScript VM often rely on
third-party native libraries written in unsafe languages for
performance.  {\erim} can isolate the runtime from bugs and
vulnerabilities in a native library by mapping the managed runtime to
{\tlib} and the native libraries to {\uapp}. This use case leverages
the ``integrity only'' version of {\erim}
(Section~\ref{sec:design:integrity-only}).  We isolated Node.js from a
native SQLite plugin. Node.js is a state-of-the-art managed runtime
for JavaScript and SQLite is a state-of-the-art database library
written in C~\cite{sqlite,nodejs}. The approach generalizes to
isolating several mutually distrusting libraries from each other by
leveraging {\erim}'s multi-component extension from
Section~\ref{sec:design:multi-threads}.

%% \todo{Reviewer B: managed
%%   runtimes typically don't only use a single native library, what
%%   about multiple distrusting libraries} \updated{While isolating only
%%   a single library in this instance, with {\erim}'s extension for
%%   multiple components described in section
%%   ~\ref{sec:design:multi-threads}, {\erim} may also be used to isolate
%%   a single trusted managed runtime from multiple native libraries.}

\paragraph*{CPI/CPS}
Code-pointer integrity (CPI)~\cite{cpi} prevents control-flow hijacks
by isolating sensitive objects---code pointers and objects that can
lead to code pointers---in a \emph{safe region} that cannot be written
without bounds checks. CPS is a lighter, less-secure variant of CPI
that isolates only code pointers.  A key challenge is to isolate the
safe region efficiently, as CPI can require switching rates on the
order of $10^6$ or more switches/s on standard benchmarks.
%
%% The original paper uses ASLR on x86-64 for its evaluation. ASLR has
%% almost no runtime overhead, but it is now known to be ineffective
%% for data
%% isolation~\cite{Shacham2004,Hund2013,Evans2015,Goktacs2016,Oikonomopoulos2016}.
We show that {\erim} can provide strong isolation for the safe region
at low cost. To do this, we override the CPI/CPS-enabled compiler's
intrinsic function for writing the sensitive region to use a call gate
around an inlined sequence of {\tlib} code that performs a bounds
check before the write. (MemSentry~\cite{Koning2017} also proposes
using MPK for isolating the safe region, but does not actually
implement it.)
%% not actually implemented according to Anjo
%We remark that this use case enforces control-flow integrity, so it is
%impossible for untrusted code to exploit a WRPKRU in {\erim}'s call
%gates and the check on lines 12--15 of the call gate can be removed
%(for this use case). However, the binary inspection and interception
%mechanism of Section~\ref{sec:binary-inspection} is still necessary to
%prevent untrusted code from mapping executable code pages with other
%occurrences of WRPKRU.

%\input{implementation}
\section{Evaluation}
\label{erim:sec:eval}

We have implemented two versions of an \erim\ prototype for
Linux.\footnote{Available online at
  \url{https://gitlab.mpi-sws.org/vahldiek/erim}.}
One version relies on a 77 line Linux Security Module (LSM) that
intercepts all mmap and mprotect calls to prevent {\uapp} from mapping
pages in executable mode, and prevents {\uapp} from overriding the
binary inspection handler. We additionally added 26 LoC for kernel
hooks to Linux v4.9.110, which were needed by the LSM.  We also
implemented {\erim} on an \emph{unmodified} Linux kernel using the
ptrace-based technique described in
Section~\ref{sec:binary-inspection}. In the following, we show results
obtained with the modified kernel. The performance of {\erim} on the
stock Linux kernel is similar, except that the costs of {\tt mmap},
{\tt mprotect}, and {\tt pkey\_mprotect} syscalls that enable execute
permissions are about 10x higher. Since the evaluated applications use
these operations infrequently, the impact on their overall performance
is negligible.

Our implementation also includes the {\erim} runtime library, which
provides a memory allocator over {\tmem}, call gates, the {\erim}
initialization code, and binary inspection. These comprise 569 LoC.
Separately, we have implemented the rewriting logic to eliminate
inadvertent WRPKRU occurrences (about 2250 LoC). While we
  have not yet integrated the logic into either a compiler or our
  inspection handler, the binaries used in our performance evaluation
  experiments do not have any unsafe WRPKRU occurrences and do not
  load any libraries at runtime. However, the binaries did have two
  legitimate occurrences of XRSTOR (in the dynamic linker library
  \texttt{ld.so}), which we made safe as described in
  Section~\ref{sec:binary-inspection}. Two other inadvertent XRSTOR
  occurred in data-only pages of executable segments in \texttt{libm},
  which is used by the SPEC benchmarks. We made these safe by
  re-mapping the pages read-only. Hence, the results we report are on
  completely safe binaries.
%that do not require further rewriting.

We evaluate the {\erim} prototype on microbenchmarks and on the three
applications mentioned in Section~\ref{sec:usecases}. Unless otherwise
mentioned, we perform our experiments on Dell PowerEdge R640 machines
with 16-core MPK-enabled Intel Xeon Gold 6142 2.6GHz CPUs (with the
latest firmware; Turbo Boost and SpeedStep were disabled), 384GB
memory, 10Gbps Ethernet links, running Debian 8, Linux kernel
v4.9.60. For the OpenSSL/webserver experiments in
Sections~\ref{erim:sec:eval:nginx} and~\ref{sec:eval:comparison}, we
use NGINX v1.12.1, OpenSSL v1.1.1 and the ECDHE-RSA-AES128-GCM-SHA256
cipher. For the managed language runtime experiment
(Section~\ref{sec:eval:managed-runtime}), we use Node.js v9.11.1 and
SQLite v3.22.0. For the CPI experiment
(Section~\ref{erim:sec:eval:cpi}), we use the Levee prototype v0.2
available from \url{http://dslab.epfl.ch/proj/cpi/} and Clang v3.3.1
including its CPI compile pass, runtime library extensions and
link-time optimization.

\subsection{Microbenchmarks}
\label{erim:sec:eval:microbenchmarks}

\begin{table}[t]
\centering
\small
\begin{tabular}{|c|c|}
\hline
Call type  & Cost (cycles)\\ \hline
Inlined call (no switch) &  5 \\ \hline
Direct call (no switch) &  8 \\ \hline
Indirect call (no switch) & 19  \\ \hline
\hline
Inlined call + switch & 60 \\ \hline
Direct call + switch &  69 \\ \hline
Indirect call + switch & 99 \\ \hline
\hline
getpid system call & 152 \\ \hline
Call + VMFUNC EPT switch & 332 \\ \hline
lwC switch~\cite{Litton2016} (Skylake CPU) & 6050\\ \hline
%SGX enter/exit~\cite{Koning2017} & 7664 & - \\ \hline \anjo{just to keep a reference to the number/taken from memsentry paper running Xeon E3 v5}
\end{tabular}
\caption{Cycle counts for basic call and return}
\label{erim:tab:microbench}
\end{table}

\paragraph{Switch cost} We performed a microbenchmark to measure the
overhead of invoking a function with and without a switch to a trusted
component.  The function adds a constant to an integer argument and
returns the result.  Table~\ref{erim:tab:microbench} shows the cost of
invoking the function, in cycles, as an inlined function (I), as a
directly called function (DC), and as a function called via a function
pointer (FP). For reference, the table also includes the cost of a
simple syscall (getpid), the cost of a switch on lwCs, a recent
isolation mechanism based on kernel page table
protections~\cite{Litton2016}, and the cost of a VMFUNC (Intel
VT-x)-based extended page table switch.

In our microbenchmark, calls with an {\erim} switch are between 55 and
80 cycles more expensive than their no-switch counterparts. The most
expensive indirect call costs less than the simplest system call
(getpid). {\erim} switches are up to 3-5x faster than VMFUNC switches
and up to 100x faster than lwC switches.

Because the CPU must not reorder loads and stores with respect to a
WRPKRU instruction, the overhead of an {\erim} switch depends on the
CPU pipeline state at the time the WRPKRUs are executed. In
experiments described later in this section, we observed average
overheads ranging from 11 to 260 cycles per switch. At a clock rate of
2.6GHz, this corresponds to overheads between 0.04\% and 1.0\% for
100,000 switches per second, which is significantly lower than the
overhead of any kernel- or hypervisor-based isolation.

%\updated{The observed WRPKRU cost are in contrast to the emulations in
%  existing work (Memsentry~\cite{Koning2017} and
%  IMIX~\cite{IMIX}). Appendix~\ref{appendix:eval-emu} shows and
%  evaluates an assembly sequence which performs like the real WRPKRU
%  instruction in experiments presented in this paper.}

\paragraph{Binary inspection} To determine the cost of {\erim}'s
binary inspection, we measured the cost of scanning the binaries of
all 18 applications in the CINT/FLOAT SPEC 2006 CPU benchmark. These
range in size from 9 to 3918 4KB pages, contain between 35 and 63765
intentional WRPKRU instructions when compiled with CPI (see
Section~\ref{erim:sec:eval:cpi}), no unintended WRPKRU and no XRSTOR
instructions. The overhead is largely independent of the number of
WRPKRU instructions and ranges between 3.5 and 6.2 microseconds per
page. Even for the largest binary, the scan takes only 17.7ms, a tiny
fraction of a typical process' runtime.

%% Appendix~\ref{appendix:staticrewriting} describes how to
%% reduce runtime costs due to binary inspection further by rewriting
%% binaries ahead of time.

%%%%%%%%%%%%%%%%%%%%%%%%%%%%%%%%%%%%%%%%%%%%%%%%%%%%%%%%%

%%%%%%%%%%%%%%%%%%%%%%%%%%%%%%%%%%%%%%%%%%%%%%%%%%%%%%%%%

\subsection{Protecting session keys in NGINX}
\label{erim:sec:eval:nginx}

\begin{figure}
  \centering
  \includegraphics[width=.95\linewidth]{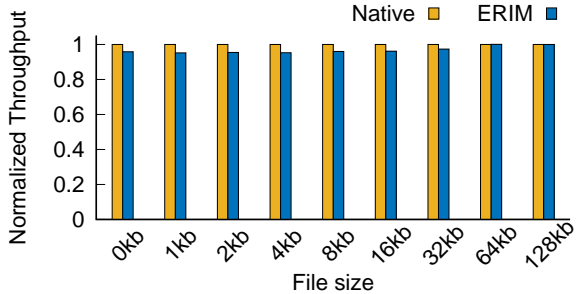}
  \caption{Throughput of NGINX with one worker, normalized to native
    (no protection), with varying request sizes. Standard deviations
    were all below 1.1\%.}
  \label{erim:fig:nginx:relative}
\end{figure}

\begin{table*}[t]
  \centering
  \small
\begin{tabular}{|c|p{1cm}|p{1.1cm}|p{1.1cm}|p{1.1cm}|p{1.1cm}|p{1.1cm}|p{1.1cm}|p{1.1cm}|}
  \hline
  \multirow{2}{1cm}{File size (KB)} & \multicolumn{2}{c|}{1 worker} & 
  \multicolumn{2}{c|}{3 workers} & \multicolumn{2}{c|}{5 workers} & \multicolumn{2}{c|}{10 workers} \\
  \cline{2-9}
  & Native (req/s) & {\erim} rel. (\%)
  & Native (req/s) & {\erim} rel. (\%)
  & Native (req/s) & {\erim} rel. (\%)
  & Native (req/s) & {\erim} rel. (\%) \\
  \hline
  0 & 95,761 & 95.8 & 276,736 & 96.1 & 466,419 & 95.7 & 823,471 & 96.4 \\
  \hline
  1 & 87,022 & 95.2 & 250,565 & 94.5 & 421,656 & 96.1 & 746,278 & 95.5 \\
  \hline
  2 & 82,137 & 95.4 & 235,820 & 95.1 & 388,926 & 96.6 & 497,778 & 100.0 \\
  \hline
  4 & 76,562 & 95.3 & 217,602 & 94.9 & 263,719 & 100.0 \\
  \cline{1-7}
  8 & 67,855 & 96.0 & 142,680 & 100.0\\
  \cline{1-5}
\end{tabular}
\caption{Nginx throughput with multiple workers. The standard
  deviation is below 1.5\% in all cases.}
\label{erim:tab:nginxresultmulti}
\end{table*}

\begin{table}[t]
\centering
\small
\begin{tabular}{|c|p{1cm}|p{1.1cm}|c|c|}
\hline
\multirow{2}{0.8cm}{File size (KB)} & \multicolumn{2}{c|}{Throughput} & \multirow{2}{*}{Switches/s} & \multirow{2}{1.3cm}{CPU load native (\%)} \\\cline{2-3}
& Native (req/s) & {\erim} rel. (\%) & &  \\ \hline
0 & 95,761 & 95.8 & 1,342,605 & 100.0 \\\hline
1 & 87,022 & 95.2 & 1,220,266 & 100.0\\\hline
2 & 82,137 & 95.4 & 1,151,877 & 100.0 \\\hline
4 & 76,562 & 95.3 & 1,073,843 & 100.0\\\hline
8 & 67,855 & 96.0 & 974,780 & 100.0 \\\hline
16 & 45,483 & 97.1 & 820,534 & 100.0 \\\hline
32 & 32,381 & 97.3 & 779,141 & 100.0 \\\hline
64 & 17,827 & 100.0 & 679,371 & 96.7 \\\hline
128 & 8,937 & 100.0 & 556,152 & 86.4\\\hline
%256 & 4474 & 4474 & 0.00 & 494,426 & 82.1 & 83.2 \\\hline
%512 & 2238 & 2238 & 0.00 & 463,754  & 80.8 & 82.4 \\\hline
%1024 & 1119 & 1119 & 0.00 & 448,954  & 83.3 & 84.5 \\\hline
%2048 & 559 & 558 & 0.11 & 441,467  & 83.1 & 84.6  \\\hline
\end{tabular}
\caption{Nginx throughput with a single worker. The standard deviation
  is below 1.1\% in all cases.}
\label{erim:tab:nginxresultsingle}
\end{table}

Next, we use {\erim} to isolate SSL session keys in a high performance
web server, NGINX.  We configured NGINX to use only the
  ECDHE-RSA-AES128-GCM-SHA256 cipher and AES encryption for sessions.
We modified OpenSSL's libcrypto to isolate all session keys and the
functions for AES key allocation and encryption/decryption into
{\erim}'s {\tlib}, and use {\erim} call gates to invoke these
functions.

%% For efficient transfer of data, the trusted component has
%% read/write access to the web server component, but not vice versa.
To measure {\erim}'s overhead on the peak throughput, we configure a
single NGINX worker pinned to a CPU core, and connect to it remotely
over HTTPS with keep-alive from 4 concurrent ApacheBench ({\tt
  ab})~\cite{ab} instances each simulating 75 concurrent clients. The
clients all request the same file, whose size we vary from 0 to 128KB
across experiments.\footnote{Since NGINX only serves static files in
  this experiment, its support for Lua and JavaScript is not used. As
  a result, this experiment does not rely on any support for Jit,
  which we have not yet implemented.}
Figure~\ref{erim:fig:nginx:relative} shows the average throughput of
10 runs of an {\erim}-protected NGINX relative to native NGINX without
any protection for different file sizes, measured after an initial
warm-up period.
%% Figure~\ref{erim:fig:nginx:reqps} shows the absolute throughputs in
%% requests/s in the same experiment.
%All numbers are averages of 10 runs.

%% \begin{figure*}[tb]
%% \centering
% \begin{figure}
%%   \centering
%%   \includegraphics[width=\linewidth]{nginx.reqps.gold1.eps}
%%   \caption{Average number of requests per second, native and
%%     {\erim}. Standard deviations were all below 1.1\%.}
%%   \label{erim:fig:nginx:reqps}
%% \end{figure}
%% \caption{Nginx throughput with one worker, with and without {\erim}
%%   protection, with varying request sizes.}

{\erim}-protected NGINX provides a throughput within 95.18\% of the
unprotected server for all request sizes. To explain the overhead
further, we list the number of {\erim} switches per second in the
NGINX worker and the worker's CPU utilization in
Table~\ref{erim:tab:nginxresultsingle} for request sizes up to
128KB. The overhead shows a general trend up to requests of size 32
KB: The worker's core remains saturated but as the request size
increases, the number of {\erim} switches per second decrease, and so
does {\erim}'s relative overhead. The observations are consistent with
an overhead of about 0.31\%--0.44\% for 100,000 switches per
second. For request sizes 64KB and higher, the 10Gbps network 
saturates and the worker does not utilize its CPU core completely in
the baseline. The free CPU cycles absorb {\erim}'s CPU overhead, so
{\erim}'s throughput matches that of the baseline.

Note that this is an extreme test case, as the web server does almost
nothing and serves the same cached file repeatedly. To get a more
realistic assessment, we set up NGINX to serve from main memory static
HTML pages from a 571 MB (15,520 pages) Wikipedia snapshot of
2006~\cite{wikiss}. File sizes vary from 417 bytes to 522 KB (average
size 37.7 KB). 75 keep-alive clients request random pages (selected
based on pageviews on Wikipedia~\cite{wikipageview}). The average
throughput with a single NGINX worker was 22,415 requests/s in the
baseline and 21,802 requests/s with {\erim} (std.\ dev.\ below 0.6\%
in both cases). On average, there were 615,000 switches a second. This
corresponds to a total overhead of 2.7\%, or about 0.43\% for 100,000
switches a second.

\paragraph*{Scaling with multiple workers}
To verify that {\erim} scales with core parallelism, we re-ran the
first experiment above with 3, 5 and 10 NGINX workers pinned to
separate cores, and sufficient numbers of concurrent clients to
saturate all the workers. Table~\ref{erim:tab:nginxresultmulti} shows
the relative overheads with different number of workers. (For requests
larger than those shown in the table, the network saturates, and
the spare CPU cycles absorb {\erim}'s overhead completely.) The
overheads were independent of the number of workers (cores),
indicating that {\erim} adds no additional synchronization and scales
perfectly with core parallelism. This result is expected as updates to
the per-core PKRU do not affect other cores.

%% (Due to lack of space, we omit the detailed results of this
%% experiment. These results are available in an online appendix, which
%% has been anonymized for review~\cite{...}. \todo{Upload appendix
%%   somewhere and fill citation for it here.})

%% \updated{Appendix~\ref{appendix:comptolwc} describes the experiment and our results in detail.}

%% (The details of this comparative experiment are also provided in
%% our online appendix.)

%%%%%%%%%%%%%%%%%%%%%%%%%%%%%%%%%%%%%%%%%%%%%

%%%%%%%%%%%%%%%%%%%%%%%%%%%%%%%%%%%%%%%%%%%%%

\subsection{Isolating managed runtimes}
\label{sec:eval:managed-runtime}

\begin{table}[t]
\centering
\small
%data based on 2018-05-02_2229
\begin{tabular}{|c|c|c|}
  \hline
Test \# & Switches/s & {\erim} overhead (\%) \\\hline
  100 & 11,183,281 & 12.73\% \\\hline
  110 & 8,329,914 & 12.18\% \\\hline
  400 & 8,161,584 & 15.42\% \\\hline
  120 & 7,190,766 & 13.81\% \\\hline
  142 & 7,074,553 & 9.41\% \\\hline
  500 & 6,419,008 & 12.13\% \\\hline
  510 & 5,868,395 & 5.60\% \\\hline
  410 & 5,091,212 & 3.64\%  \\\hline
  240 & 2,358,524 & 3.74\%  \\\hline
  280 & 2,303,516 & 3.22\%  \\\hline
  170 & 1,264,366 & 4.22\%  \\\hline
  310 & 1,133,364 & 2.92\% \\\hline
  161 & 1,019,138 & 2.81\%  \\\hline
  160 & 1,014,829 & 2.73\% \\\hline
  230 & 670,196 & 2.04\% \\\hline
  270 & 560,257 & 2.28\%  \\\hline

\end{tabular}
\caption{Overhead relative to native execution for SQLite speedtest1
  tests with more than 100,000 switches/s. Standard deviations were
  below 5.6\%.
%  for native and {\erim}.  and below 15.4\% for  WebAssembly.
}
\label{erim:tab:sqlite}
\end{table}

Next, we use {\erim} to isolate a managed language runtime from an
untrusted native library. Specifically, we link the widely-used C
database library, SQLite, to Node.js, a state-of-the-art JavaScript
runtime and
%use {\erim} to isolate Node.js from SQLite by mapping
map Node.js's runtime to {\tlib} and SQLite to {\uapp}. We modified
SQLite's entry points to invoke call gates.  To isolate Node.js's
stack from SQLite, we run Node.js on a separate stack in {\tmem}, and
switch to the standard stack (in {\umem}) prior to calling a SQLite
function. Finally, SQLite uses the libc function {\tt memmove}, which
accesses libc constants that are in {\tmem}, so we implemented a
separate {\tt memmove} for SQLite. In total, we added 437 LoC.%
%
%% \footnote{Our implementation of the isolation isn't yet
%%   complete. Specifically, we currently lack the code to
%%   ... However, our instrumentation of the call gates is complete,
%%   so the overheads reported here are exactly what they will be with
%%   the full isolation in place.}
%

We measure overheads on the speedtest1 benchmark that comes with
SQLite and emulates a typical database workload~\cite{speedtest1}. The
benchmark performs 32 short tests that stress different database
functions like selects, joins, inserts and deletes. We increased the
iterations in each test by a factor of four to make the tests
longer. Our baseline for comparison is native SQLite linked to
Node.js without any protection. We configure the benchmark to store
the database in memory and report averages of 20 runs.

The geometric mean of {\erim}'s runtime overhead across all tests is
4.3\%. The overhead is below 6.7\% on all tests except those with more
than $10^6$ switches per second. This suggests that {\erim} can be
used for isolating native libraries from managed language runtimes
with low overheads up to a switching cost of the order of $10^6$ per
second. Beyond that the overhead is
noticeable. Table~\ref{erim:tab:sqlite} shows the relative overheads
for tests with switching rates of at least 100,000/s. The numbers are
consistent with an average overhead between 0.07\% and 0.41\% for
100,000 switches/s. The actual switch cost measured from direct CPU
cycle counts varies from 73 to 260 cycles across all tests. It exceeds
100 cycles only when the switch rate is less than 2,000 times/s. We
verified that these are due to i-cache misses---at low switch rates,
the call gate instructions are evicted between switches.

%% All other tests have average switch rates below
%% \todos{2,000}/s. Extrapolating the switch rate to 100,000 switches/s
%% from these tests is meaningless since that also magnifies the
%% experimental error (standard deviations) proportionally.

%%%%%%%%%%%%%%%%%%%%%%%%%%%%%%%%%%%%%%%%%%%%%%%%%%%%%%%%%%%%%%%%%%%%%%

%%%%%%%%%%%%%%%%%%%%%%%%%%%%%%%%%%%%%%%%%%%%%%%%%%%%%%%%%%%%%%%%%%%%%%

\subsection{Protecting sensitive data in CPI/CPS}
\label{erim:sec:eval:cpi}

\begin{figure}[t]
  \centering
  \includegraphics[width=.95\linewidth, angle=270]{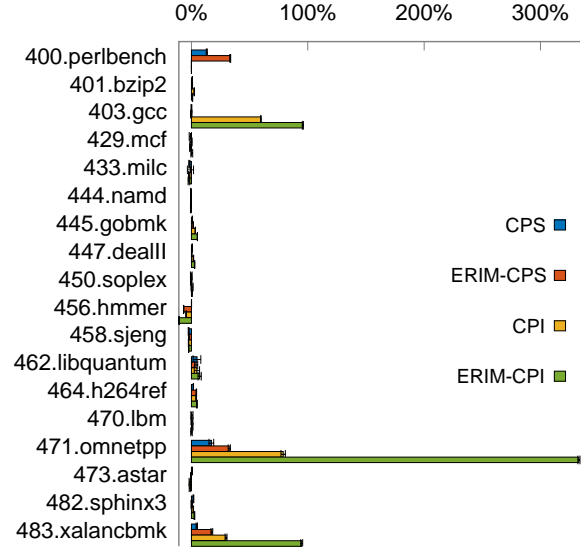}
  \caption{Percentage overhead relative to no protection.}
    %% \anjo{\href{https://docs.google.com/spreadsheets/d/1xN9FOYwGGQ_S0Mg9PAxUwMdaS9g34ahmgVZ58abQ6hg/edit?usp=sharing}{Raw data}}}
  \label{erim:fig:spec:reloverheadnativ}
\end{figure}

Next, we use {\erim} to isolate the safe region of CPI and
CPS~\cite{cpi} in a separate domain. We modified CPI/CPS's LLVM
compiler pass to emit additional {\erim} switches, which bracket any
code that modifies the safe region. The switch code, as well as the
instructions modifying the safe region, are inlined with the
application code.
In addition, we implemented simple optimizations to safely reduce the
frequency of {\erim} domain switches. For instance, the original
implementation sets sensitive code pointers to zero during
initialization.  Rather than generate a domain switch for each pointer
initialization, we generate loops of pointer set operations that are
bracketed by a single pair of \erim\ domain switches.  This is safe
because the loop relies on direct jumps and the code to set a pointer
is inlined in the loop's body. In all, we modified 300 LoC in LLVM's
CPI/CPS pass.

Like the original CPI/CPS paper~\cite{cpi}, we compare the overhead of
the original and our \erim-protected CPI/CPS system on the SPEC CPU
2006 CINT/FLOAT benchmarks, relative to a baseline compiled with Clang
without any protection. The original CPI/CPS system is configured to
use ASLR for isolation, the default technique used on x86-64 in the
original paper. ASLR imposes almost no switching overhead, but also
provides no
security~\cite{Shacham2004,Hund2013,Evans2015,Goktacs2016,Oikonomopoulos2016}.

\if 0
Figure~\ref{erim:fig:spec:reloverheadnativ} shows the average runtime
overhead over 10 runs of the original CPI/CPS (lines ``CPI/CPS'')and
CPI/CPS with {\erim} (lines ``{\erim}-CPI/CPS''). For comparison, we
also show the overhead of CPI/CPS on {\erim} without binary inspection
(lines ``{\erim}-CPI/CPS without scan''). All overheads are normalized
to the unprotected SPEC benchmark. We were unable to obtain results
for 400.perlbench for CPI and 453.povray for both CPS and CPI. The
400.perlbench benchmark does not halt when compiled with CPI and the
SPEC internal result verification for 453.povray fails due to
unexpected output. These problems exist in the code generated by the
Levee CPI/CPS prototype with CPI/CPS enabled (\mbox{-fcps/-fcpi}), not
our modifications.
\fi

Figure~\ref{erim:fig:spec:reloverheadnativ} shows the average runtime
overhead of 10 runs of the original CPI/CPS (lines ``CPI/CPS'') and
CPI/CPS over {\erim} (lines ``{\erim}-CPI/CPS''). All overheads are
normalized to the unprotected SPEC benchmark. We could not obtain
results for 400.perlbench for CPI and 453.povray for both CPS and
CPI. 400.perlbench does not halt when compiled with CPI
and SPEC's result verification for 453.povray fails due to unexpected
output. These problems exist in the code generated by the Levee
CPI/CPS prototype with CPI/CPS enabled (\mbox{-fcps/-fcpi}), not our
modifications.

%% As expected, {\erim}'s binary inspection adds neglible overhead (the
%% difference in {\erim}'s overheads with and without binary inspection
%% is less than \todos{XXX}\% on all benchmarks). In the following, we
%% discuss only {\erim} \emph{without} binary inspection, since this
%% allows us to quantify the switching cost.

%Figure~\ref{erim:fig:spec:reloverheadcpscpi} shows the relative
%difference between CPS/CPI and its \erim-based counter part.
%% This shows that \erim\ can lend
%% hardware-strength isolation to CPS on x86-64 at low cost.

\begin{table}[t]
  \centering
  \small
\begin{tabular}{|l|c|c|}
\hline
\multirow{2}{*}{Benchmark} & \multirow{2}{*}{Switches/sec} & {\erim}-CPI overhead \\
& & relative to\ orig.\ CPI in \%\\ \hline
%401.bzip2 & 1 & -1.75\%\\\hline 
403.gcc & 16,454,595 & 22.30\%\\\hline 
%429.mcf & 2,339 & 1.81\%\\\hline 
%433.milc & 0 & 0.49\%\\\hline 
%444.namd & 14 & -0.16\%\\\hline 
445.gobmk & 1,074,716 & 1.77\%\\\hline 
447.dealII & 1,277,645 & 0.56\%\\\hline 
450.soplex & 410,649 & 0.60\%\\\hline 
%456.hmmer & 5,965 & -0.05\%\\\hline 
%458.sjeng & 0 & 0.32\%\\\hline 
%462.libquantum & 23,327 & 1.38\%\\\hline 
464.h264ref & 1,705,131 & 1.22\%\\\hline 
%470.lbm & 0 & -0.23\%\\\hline 
471.omnetpp & 89,260,024 & 144.02\%\\\hline 
%473.astar & 38,508 & 0.71\%\\\hline 
482.sphinx3 & 1,158,495 & 0.84\%\\\hline 
483.xalancbmk & 32,650,497 & 52.22\%\\\hline 

% data till usenix sec
%401.bzip2     & 1.8  & --1.5 \\ \hline
%403.gcc       & 13,454,647 & 22.3 \\ \hline
%429.mcf       & 2069 & 3.3 \\ \hline
%433.milc      & 0.5 & -0.36 \ \hline
%444.namd      & 19.3 & -.12 \\ \hline
%445.gobmk     & 1,055,994 & 1.77 \\ \hline
%447.dealII    & 1,270,582 & 0.56 \\ \hline
%450.soplex    & 408,192 & 0.6 \\ \hline
%456.hmmer     & 8222 & -0.65 \\ \hline
%458.sjeng     & 0.1 & -.6 \\ \hline
%462.libquantum & 37,383 & 0 \\ \hline
%464.h264ref   & 1,684,572 & 1.22 \\ \hline
%470.lbm       & 0.1 & 0.24    \\ \hline
%471.omnetpp   & 36,578,718 & 144.02 \\ \hline
%473.astar     & 46,830 & -.63 \\ \hline
%482.sphinx    & 1,148,883 &  0.84\\ \hline
%483.xalancbmk  & 21,448,977  & 52.22\\ \hline
\end{tabular}
\caption{Domain switch rates of selected SPEC CPU benchmarks and
  overheads for {\erim}-CPI without binary inspection, \emph{relative
    to the original CPI with ASLR}.}
\label{tab:spec-switchcounts}
\end{table}

\emph{CPI}: The geometric means of the overheads (relative to no
protection) of the original CPI and {\erim}-CPI across all benchmarks
are 4.7\% and 5.3\%, respectively. The relative overheads of
{\erim}-CPI are low on all individual benchmarks except gcc, omnetpp,
and xalancbmk.

To understand this better, we examined switching rates across
benchmarks. Table~\ref{tab:spec-switchcounts} shows the switching
rates for benchmarks that require more than 100,000 switches/s. From
the table, we see that the high overheads on gcc, omnetpp and
xalancbmk are due to extremely high switching rates on these three
benchmarks (between $1.6\times10^7$ and $8.9\times10^7$ per second).
Further profiling indicated that the reason for the high switch rate
is tight loops with pointer updates (each pointer update incurs a
switch). An optimization pass could hoist the domain switches out of
the loops safely using only direct control flow instructions and
enforcing store instructions to be bound to the application memory,
but we have not implemented it yet.

Table~\ref{tab:spec-switchcounts} also shows the overhead of
{\erim}-CPI excluding binary inspection, relative to the original CPI
over ASLR (not relative to an unprotected baseline as in
Figure~\ref{erim:fig:spec:reloverheadnativ}). This relative overhead
is exactly the cost of {\erim}'s switching.  Depending on the
benchmark, it varies from 0.03\% to 0.16\% for 100,000 switches per
second or, equivalently, 7.8 to 41.6 cycles per switch. These results
again indicate that \erim\ can support inlined reference monitors with
switching rates of up to $10^6$ times a second with low
overhead. Beyond this rate, the overhead becomes noticeable.

\emph{CPS}: The results for CPS are similar to those for CPI, but the
overheads are generally lower. Relative to the baseline without
protection, the geometric means of the overheads of the original CPS
and {\erim}-CPS are 1.1\% and 2.4\%, respectively. {\erim}-CPS's
overhead relative to the original CPS is within 2.5\% on all
benchmarks, except except perlbench, omnetpp and xalancbmk, where it
ranges up to 17.9\%.

%%%%%%%%%%%%%%%%%%%%%%%%%%%%%%%%%%%%%%%%%%%%%%%%%%%%%%%%%%%%%%%%%%%%%%%

%%%%%%%%%%%%%%%%%%%%%%%%%%%%%%%%%%%%%%%%%%%%%%%%%%%%%%%%%%%%%%%%%%%%%%%

\subsection{Comparison to existing techniques}
\label{sec:eval:comparison}

%% In this section we compare {\erim}'s single-worker NGINX
%% performance against an MPX-based instrumentation preventing
%% untrusted memory accesses from accessing trusted memory, a VMFUNC
%% extended page table switch isolating the session keys, and a
%% kernel-based isolation of session
%% keys. Figure~\ref{erim:fig:nginx:compare} summarizes the results.

{In this section}, we compare {\erim} to isolation using SFI (with
Intel MPX), extended page tables (with Intel VT-x/VMFUNC), kernel page
tables (with lwCs), and instrumentation of untrusted code for full
memory safety (with WebAssembly). In each case, our primary goal is a
\emph{quantitative} comparison of the technique's overhead to that of
     {\erim}. As we show below, {\erim}'s overheads are substantially
     lower than those of the other techniques. But before presenting
     these results, we provide a brief \emph{qualitative} comparison
     of the techniques in terms of their threat models.

\paragraph{Qualitative comparison of techniques}
Isolation using standard kernel page tables affords a threat model
similar to {\erim}'s. In particular, like {\erim}, the OS kernel must
be trusted. In principle, isolation using a hypervisor's
  extended page tables (VMFUNC) can afford a stronger threat model, in
  which the OS kernel need not be trusted~\cite{Liu2015}.

Isolation using SFI, with or without Intel MPX, affords a threat model
weaker than {\erim}'s since one must additionally trust the transform
that adds bounds checks to the untrusted code. For full protection, a
control-flow integrity (CFI) mechanism is also needed to prevent
circumvention of bounds checks. This further increases both the
trusted computing base (TCB) and the overheads. In the experiments
below, we omit the CFI defense, thus underestimating SFI overheads for
protection comparable to {\erim}'s.

Instrumenting untrusted code for full memory safety, i.e.,
bounds-checking at the granularity of individual memory allocations,
implicitly affords the protection that SFI provides. Additionally,
such instrumentation also protects the untrusted code's data from
other outside threats, a use case that the other techniques here
(including {\erim}) do not handle. However, as for SFI, the mechanism
used to instrument the untrusted code must be trusted. In our
experiments below, we enforce memory safety by compiling untrusted
code to Web\-Assembly, and this compiler must be trusted.

%% Isolation afforded by full memory safety in untrusted
%% code has a threat model incomparable to that of {\erim}. The mechanism
%% that instruments the untrusted code for memory safety must be trusted
%% (as in SFI), but full memory safety can protect the untrusted code's
%% data from other outside threats, 

Next, we quantitatively compare the overheads of these techniques
to those of {\erim}.

%% We compare the overhead of {\erim} to modern isolation techniques
%% using SFI (with Intel MPX), extended page tables (with Intel
%% VT-x/VMFUNC), kernel page tables (with lwCs), and instrumentation
%% for memory-safety (with WebAssembly). For all but the last
%% comparison, we use the NGINX experiment of
%% Section~\ref{erim:sec:eval:nginx}.

\begin{figure}[t!]
  \centering
\begin{subfigure}[t]{0.95\linewidth}
  \centering
  \includegraphics[width=.95\linewidth]{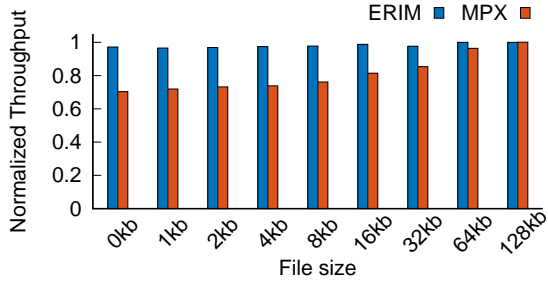}
  \caption{ERIM vs.\ SFI using MPX (averages of 3 runs, std.\ devs.\ below 1.9\%)}
  \label{erim:fig:nginx:mpx}
\end{subfigure}
\\\vspace{0.2cm}
\begin{subfigure}[t]{0.95\linewidth}
  \centering
  \includegraphics[width=.95\linewidth]{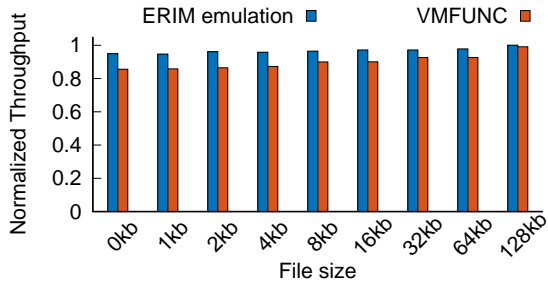}
  \caption{Emulated ERIM vs.\ VMFUNC (averages of 3 runs,
    std.\ devs.\ below 0.9\%)}
  \label{erim:fig:nginx:vmfunc}
\end{subfigure}
\\\vspace{0.2cm}
\begin{subfigure}[t]{0.95\linewidth}
  \centering
  \includegraphics[width=.95\linewidth]{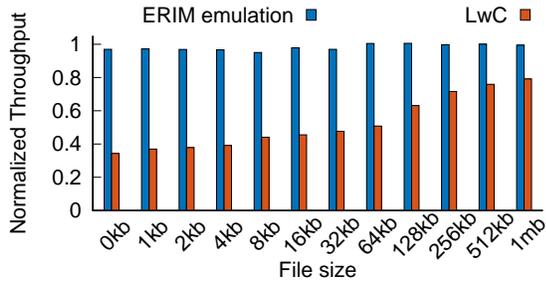}
  \caption{Emulated ERIM vs.\ LwC (averages of 5 runs,
    std.\ devs.\ below 1.1\%)}
  \label{erim:fig:nginx:lwcs}
\end{subfigure}
\caption{Comparison of NGINX throughput with {\erim} and alternative
  isolation techniques}
\end{figure}

\paragraph{SFI using MPX}
We start by comparing the cost of {\erim}'s isolation to that of
isolation based on SFI using MPX. For this, we follow the NGINX
experiment of Section~\ref{erim:sec:eval:nginx}. We place OpenSSL
(trusted) in a designated memory region, and use
MemSentry~\cite{Koning2017} to compile all of NGINX (untrusted) with
MPX-based memory-bounds checks that prevent it from accessing the
OpenSSL region directly.%
\footnote{This setup \emph{reduces} the overheads of MPX as compared
  to the setup of Section~\ref{erim:sec:eval:nginx}, which isolates
  only small parts of OpenSSL. It is also less secure. Hence, the MPX
  overheads reported here are conservative.}
To get comparable measurements on the (no protection) baseline and
{\erim}, we recompile NGINX with Clang version 3.8, which is the
version that MemSentry supports. We then re-run the single worker
experiments of Section~\ref{erim:sec:eval:nginx}.

Figure~\ref{erim:fig:nginx:mpx} shows the overheads of MPX and {\erim}
on NGINX's throughput, relative to a no-protection baseline. The
MPX-based instrumentation reduces the throughput of NGINX by 15-30\%
until the experiment is no longer CPU-bound (file sizes $\geq$
64kb). In contrast, {\erim} reduces overheads by no more than
3.5\%. Across all file sizes, MPX overheads are 4.2-8.5x those of
{\erim}.

MPX (more generally, SFI) and {\erim} impose overhead in different
ways. MPX imposes an overhead during the execution of NGINX (the
untrusted component), while {\erim} imposes an overhead on component
switches. Consequently, one could argue that, as the switch rate
increases, {\erim} \emph{must} eventually become more expensive than
MPX. While this is theoretically true, in this experiment, we already
observe extremely high switch rates of 1.2M/s (for file size 0kb) and,
even then, MPX's overhead is 8.4x that of {\erim}'s overhead.

Further, as explained earlier, for strong security, SFI must be
supported by control-flow integrity, which would induce additional
overheads that are not included here.

%% Finally, despite its higher overhead, the MPX-based technique is less
%% secure than {\erim} because the former does not defend against
%% control-flow attacks that bypass bounds checks. To prevent such
%% attacks, a separate control-flow integrity technique must be deployed,
%% which would have additional overheads.

%% We use MemSentry~\cite{Koning2017} instrumentation to
%% prevent untrusted memory accesses from accessing trusted memory by
%% inserting Intel MPX-based bound checks. We use their LLVM compiler
%% pass to instrument all of NGINX, but not OpenSSL treating all of
%% OpenSSL as trusted. This is conservative compared to {\erim}, since
%% for {\erim} we single out only functions handeling the session
%% keys. As a result, bugs and vulnerabilities in OpenSSL may be
%% exploited, whereas {\erim} protects against bugs and vulnerabilities
%% in large parts of OpenSSL. To match the compiler of MemSentry, native
%% and {\erim} NGINX are compiled with Clang 3.8 and experiments are
%% repeated 3 times (std. dev. is blow
%% 1.9\%). 

%% MemSentry's MPX-based instrumentation reduces the throughput of
%% NGINX by 15-30\% until the native NGINX experiment is no longer
%% CPU bound. The available CPU cycles are selvaged for bounds checks
%% increasing the CPU utilization between 10-20\%.

%% Instead, {\erim} reduces the throughput only by upto 3.5\% and the CPU
%% utilization only increases by upto 2\%. This is 4.2-8.5x less throughput
%% reduction.

\paragraph{Extended page tables (VMFUNC)}
Next, we compare {\erim} to isolation based on extended page tables
(EPTs) using Intel VT-x and VMFUNC. To get access to EPTs, we use
Dune~\cite{Belay2012} and a patch from MemSentry. We create two page
tables---one maps the trusted region that contains session keys, and
the other maps the untrusted region that contains all the remaining
state of NGINX and OpenSSL. Access to the first table is efficiently
switched on or off using the VMFUNC EPT switch call provided by the
MemSentry patch. This call is faster than an OS process switch since
it does not switch the process context or registers. Since we use
Dune, the OS kernel runs in hypervisor mode. It has the switch
overheads of hypervisor-based isolation using VMFUNC but includes the
OS kernel in the TCB.

Unfortunately, MemSentry's patch works only on old Linux kernels
which do \emph{not} have the page table support needed for MPKs and,
hence, cannot support {\erim}. Consequently, for this comparison, we
rely on an emulation of {\erim}'s switch overhead using standard x86
instructions. This emulation is described later in this section,
and we validate that it is accurate to within 2\% of {\erim}'s actual
overheads on a variety of programs. So we believe that the comparative
results presented here are quite accurate.

Figure~\ref{erim:fig:nginx:vmfunc} shows the throughput of NGINX
protected with VMFUNC and emulated {\erim}, relative to a baseline
with no protection for different file sizes (we use Linux kernel
v3.16). Briefly, VMFUNC induces an overhead of 7-15\%, while the
corresponding overhead of emulated {\erim} is 2.1-5.3\%. Because
both VMFUNC and {\erim} incur overhead on switches, overheads of both
reduce as the switching rate reduces, which happens as the file size
increases. (The use of Dune and extended page tables also induces an
overhead on all syscalls and page walks in the VMFUNC isolation.)

To directly compare VMFUNC's overheads to \emph{actual} {\erim}'s, we
calculated VMFUNC's overhead as a function of switch rate. Across
different file sizes, this varies from 1.4\%-1.87\% for 100,000
switches/s. In contrast, actual {\erim}'s overhead in the similar
experiment of Section~\ref{erim:sec:eval:nginx} never exceeds 0.44\%
for 100,000 switches/s. This difference is consistent with the
microbenchmark results in Table~\ref{erim:tab:microbench}.

\paragraph{Kernel page tables (lwCs)}
Next, we compare {\erim}'s overhead to that of lwCs~\cite{Litton2016},
a recent system for in-process isolation based on kernel page-table
protections. LwCs map each isolated component to a separate address
space in the same process. A switch between components requires kernel
mediation to change page tables, but does not require a process
context switch. To measure lwC overheads, we re-run the NGINX
experiment of Section~\ref{erim:sec:eval:nginx}, using two lwC
contexts, one for the session keys and encryption/decryption functions
and the other for NGINX and the rest of OpenSSL. Unfortunately, lwCs
were prototyped in FreeBSD, which does not support MPK, so we again
use our emulation of {\erim}'s switch overhead to compare. All
experiments reported here were run on Dell OptiPlex 7040 machines with
4-core Intel Skylake i5-6500 CPUs clocked at 3.2 GHz, 16 GB memory, 10
Gbps Ethernet cards, and FreeBSD 11.

Figure~\ref{erim:fig:nginx:lwcs} shows the throughput of NGINX running
with lwCs and emulated {\erim}, relative to a baseline without any
protection. With lwCs, the throughput is never above 80\% of the
baseline, and for small files, where the switch rate is high, the
throughput is below 50\%. In contrast, the throughput with emulated
{\erim} is within 95\% of the baseline for all file sizes.
In terms of switch rates, lwCs incur a cost of 10.5-18.3\% for 100,000
switches/s across different file sizes. \emph{Actual} {\erim}'s switch
overhead during the similar experiment of
Section~\ref{erim:sec:eval:nginx} is no more than 0.44\% across all
file sizes, which is two orders of magnitude lower than that of lwCs.

%% Since lwC is implemented in FreeBSD without Intel MPK support, these
%% experiments were performed on Dell OptiPlex 7040 machines with 4-core
%% Intel Skylake i5-6500 CPUs clocked at 3.2 GHz, 16 GB memory, 10 Gbps
%% ehternet cards, running FreeBSD 11.

%% The throughput of NGINX with lwC-based isolation is never above 80\%
%% of native NGINX even for large files (1 mb) and, for small requests,
%% where the switch rate is higher, it is below 50\% of native NGINX. In
%% contrast, with {\erim}'s isolation, the throughput is within 95\% of
%% native NGINX in all configurations.  Hence, {\erim} performs
%% significantly better than kernel-mediated isolation.  The cost of
%% 100,000 switches/s for lwC-based isolation is between 10.5-18.3\% in
%% this experiment.

\paragraph{Memory safety (WebAssembly)}
Finally, we compare {\erim}'s overheads to those of full memory safety
on untrusted code. Specifically, we compare to compilation of
untrusted code through
WebAssembly~\cite{DBLP:conf/pldi/HaasRSTHGWZB17}, a memory-safe,
low-level language that is now supported natively by all major web
browsers and expected to replace existing SFI techniques like
  Native Client in the Chrome web browser.
%
%% Note that full memory safety is an overkill forcoarse-grained
%% isolation. Nonetheless, we make this comparison since WebAssembly is
%% expected to replace existing SFI techniques like Native Client in web
%% browsers such as Chrome.
%
We compare to {\erim} using the experiment of
Section~\ref{sec:eval:managed-runtime}. We re-compile the (untrusted)
SQLite library to WebAssembly via emscripten v1.37.37's WebAssembly
backend~\cite{emscripten}, and run the WebAssembly within Node.js,
which supports the language. Accross tests of
Table~\ref{erim:tab:sqlite}, the overhead of using WebAssembly varies
from 81\% to 193\%, which is one to two orders of magnitude higher
than {\erim}'s overhead.

\paragraph{Emulating {\erim}'s switch cost}
\label{erim:sec:eval-emu}

We describe how we emulate {\erim}'s switch cost when comparing to
VMFUNC and lwCs above. Specifically, we need to emulate the cost of a
WRPKRU instruction, which isn't natively supported in the environments
of those experiments. We do this using xor instructions to consume the
appropriate number of CPU cycles, followed by RDTSCP, which causes a
pipeline stall and prevents instruction re-ordering. Specifically, we
execute a loop five times, with
\texttt{xor eax,ecx; xor ecx,eax; xor eax,ecx}, followed by a single RDTSCP after the loop.
\if 0
We show the exact emulating sequence below.
%% \begin{lstlisting} [label={erim:lst:emulatewrpkru:rdtscp},showlines=false, caption={WRPKRU emulation using RDTSCP and XorSwitch}]
\noindent
\mbox{}~~\texttt{for(i = 0; i < 5; i++) \{}\\
\mbox{}~~~~~\texttt{xor eax,ecx};\\
\mbox{}~~~~~\texttt{xor ecx,eax};\\
\mbox{}~~~~~\texttt{xor eax,ecx};\\
\mbox{}~~\}\\
\mbox{}~~\texttt{rdtscp}
%\end{lstlisting}
\fi

To validate the emulation
%closely match those of the actual WRPKRU instruction,
we re-ran the SPEC CPU 2006 benchmark with CPI/CPS
(Section~\ref{erim:sec:eval:cpi}) after swapping actual WRPKRU
instructions with the emulation sequence shown above and compared the
resulting overheads. In each \emph{individual} test, the difference in
overhead between actual {\erim} and the emulation is below 2\%.
%This suggests that the emulation is quite accurate.
We note that a perfectly precise emulation is impossible since
emulation cannot exactly reproduce the effects of WRPKRU on the
execution pipeline. (WRPKRU must prevent the reordering of loads and
stores with respect to itself.) Depending on the specific benchmark,
our emulation slightly over- or under-estimates the actual performance
impact of WRPKRU.
%due to the reason mentioned above.
We also observed that emulations of WRPKRU using LFENCE or MFENCE (the
latter was suggested by~\cite{Koning2017}) in place of RDTSCP incur
too little or too much overhead, respectively.

\section{Conclusion}
\label{erim:sec:conclusion}

%% We conclude the paper with a brief summary of {\erim} and how it
%% compares to other memory isolation techniques.
Relying on the recent Intel MPK ISA extension and simple binary
inspection, {\erim} provides hardware-enforced isolation with an
overhead of less than 1\% for every 100,000 switches/s between
components on current Intel CPUs, and almost no overhead on execution
within a component. {\erim}'s switch cost is up to two orders of
magnitude lower than that of kernel page-table based isolation, and up
to 3-5x lower than that of VMFUNC-based isolation. For VMFUNC,
virtualization can cause additional overhead on syscalls and page
table walks. {\erim}'s overall overhead is lower than that of
isolation based on memory-bounds checks (with Intel MPX), even at
switch rates of the order of $10^6$/s. Additionally, such techniques
require control-flow integrity to provide strong security, which has
its own overhead.  {\erim}'s comparative advantage prominently stands
out on applications that switch very rapidly and spend a nontrivial
fraction of time in untrusted code.

\paragraph{Acknowledgements}
We thank the anonymous reviewers, our shepherd Tom Ritter, Bobby
Bhattacharjee, and Mathias Payer for their feedback, which helped
improve this paper. This work was supported in part by the European
Research Council (ERC Synergy imPACT 610150) and the German Science
Foundation (DFG CRC 1223).

{%\normalsize
  \bibliographystyle{plain}
  \bibliography{erim}}

%\newpage

%\appendix

%\input{appendixB}

%\input{revisioncomments}

\end{document}